%% file: main.tex
\documentclass[sn-nature]{sn-jnl}
\usepackage{graphicx}%
\usepackage{multirow}%
\usepackage{amsmath,amssymb,amsfonts}%
\usepackage{amsthm}%
\usepackage{mathrsfs}%
\usepackage{comment}%
\usepackage[title]{appendix}%
\usepackage{xcolor}%

\usepackage{textcomp}%
\usepackage{manyfoot}%
\usepackage{booktabs}%
\usepackage{algorithm}%
\usepackage{algorithmicx}%
\usepackage{algpseudocode}%
\usepackage{listings}%
\usepackage{longtable}%
\usepackage{tabularx}%
\usepackage{adjustbox}%
\usepackage{float}%
\usepackage[section]{placeins}
\usepackage{lmodern}%
\usepackage{mathrsfs}%
\usepackage{bbold}
\usepackage{siunitx}
\graphicspath{ {./images/} }

\usepackage{xspace}

\begin{document}
\newcommand{\bmfmrna}{\texttt{BMFM-RNA}}
\newcommand{\scGPT}{\texttt{scGPT}}
\newcommand{\scBERT}{\texttt{scBERT}}
\newcommand{\BiomedMultiOmic}{\texttt{Biomed-Multi-Omic}}
\newcommand{\biomedmultiomic}{\texttt{biomed-multi-omic}}
\newcommand{\scFoundation}{\texttt{scFoundation}}
\newcommand{\scPrint}{\texttt{scPrint}}
\newcommand{\Geneformer}{\texttt{Geneformer}}
\newcommand{\esmTwo}{\texttt{ESM2}}

\newcommand{\DNABERTTwo}{\texttt{DNABERT2}}
\newcommand{\BERT}{\texttt{BERT}}
\newcommand{\ModernBERT}{\texttt{ModernBERT}}
\newcommand{\Performer}{\texttt{Performer}}
\newcommand{\Nystromformer}{\texttt{Nyströmformer}}

\newcommand{\scEval}{\texttt{scEval}}
\newcommand{\UCE}{\texttt{Universal Cell Embeddings}}

\newcommand{\cellxgene}{CELLxGENE}
\newcommand{\PanglaoDB}{PanglaoDB}

\title[Article Title]{BMFM-RNA: whole-cell expression decoding improves transcriptomic foundation models}


\author[1]{{Michael M.} {Danziger}}
\author[2]{{Bharath} {Dandala}}
\author[1]{{Viatcheslav} {Gurev}}
\author[1]{{Matthew} {Madgwick}}
\author[1]{{Sivan} {Ravid}}
\author[1]{{Tim} {Rumbell}}
\author[1]{{Akira} {Koseki}}
\author[1]{{Tal} {Kozlovski}}
\author[1]{{Ching-Huei} {Tsou}}
\author[1]{{Ella} {Barkan}}
\author[1]{{Tanwi} {Biswas}}
\author[3,4]{{Jielin} {Xu}}
\author[1]{{Yishai} {Shimoni}}
\author[1]{{Jianying} {Hu}}
\author[1]{{Michal} {Rosen-Zvi}}

\affil[1]{\orgdiv{IBM Research}, \orgname{IBM Research}}
\affil[2]{\orgname{Autonomize.ai}}
\affil[3]{\orgdiv{Cleveland Clinic Genome Center, Lerner Research Institute}, \orgname{Cleveland Clinic, Cleveland, Ohio, USA}}
\affil[4]{\orgdiv{Genomic Medicine Institute, Lerner Research Institute}, \orgname{Cleveland Clinic, Cleveland, Ohio, USA}}

\abstract{Transcriptomic foundation models pretrained with masked language modeling can achieve low pretraining loss yet produce poor cell representations for downstream tasks. 
We introduce whole-cell expression decoding (WCED), where models reconstruct the entire gene vocabulary from a single CLS token embedding, even with limited inputs, creating a maximally informative bottleneck. 
WCED consistently outperforms MLM on all downstream metrics despite higher reconstruction error during training. 
Gene-level error tracking reveals that both methods preferentially learn genes whose expression co-varies with stable transcriptional programs rather than those driven by transient factors. 
We further add hierarchical cross-entropy loss that exploits Cell Ontology structure for zero-shot annotation at multiple granularity levels. 
Models trained with these objectives achieve best overall performance across CZI benchmarks, on zero-shot batch integration and linear probing cell-type annotation. 
Methods are implemented in \biomedmultiomic (\url{https://github.com/BiomedSciAI/biomed-multi-omic}), an open-source framework for transcriptomic foundation model development.
}

\maketitle

\section{Main} 

Single-cell profiling has transformed our ability to resolve cellular heterogeneity, and community-curated atlases such as CELLxGENE \cite{program_cz_2025} and the Human Cell Atlas \cite{regev_human_2017} now aggregate hundreds of millions of cells spanning diverse tissues, species and assay modalities.
Transcriptomic foundation models (TFMs) aim to exploit these resources by learning compact latent representations of single-cell RNA sequencing (scRNA-seq) data that capture biological cell state while remaining invariant to technical noise, and be efficiently adapted to downstream tasks via fine-tuning.\cite{yang_scbert_2022, theodoris_transfer_2023, cui_scgpt_2024, hao_scfoundation_2024, zeng_cellfm_2025, kalfon_scprint_2025}.
Masked language modeling (MLM), the dominant TFM pretraining paradigm, is a natural fit compared to autoregressive models for expression data because gene sets lack any inherent ordering and permutation-invariant masking provides an appropriate inductive bias~\cite{devlin2018bert}.
However, MLM can succeed at its training objective without necessarily compressing a global cell state, achieving low pretraining loss with poor downstream embeddings.

\begin{figure*}[t]
  \centering
  \includegraphics[width=0.97\textwidth]{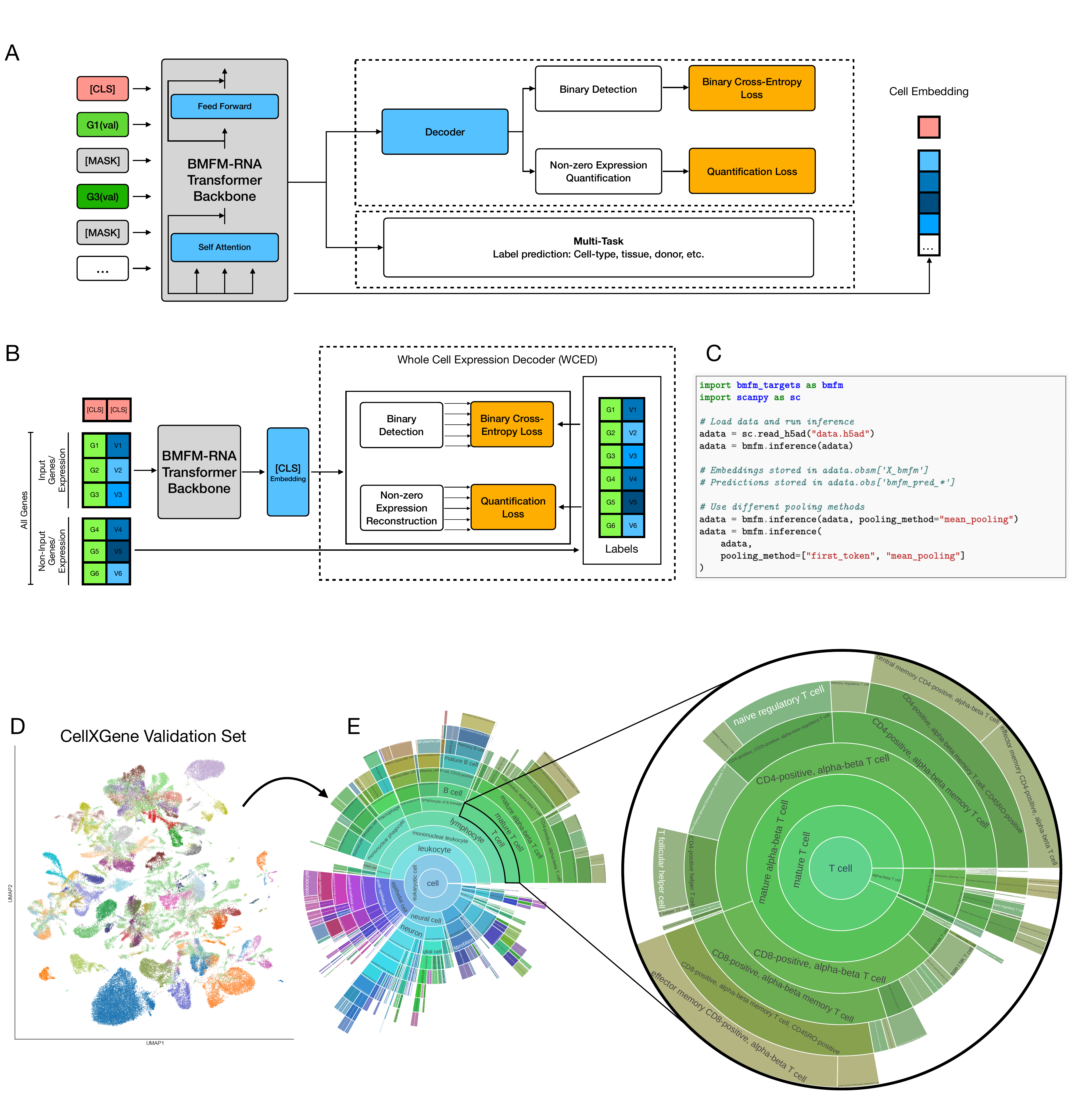}
  \caption{\textbf{Pretraining overview of BMFM-RNA models.}
  \textbf{A.} Multitask architecture design.
  \textbf{B.} Whole-cell expression decoder (WCED) training objective. Part of the cell's expression data is input into the model (green box) while some of it is left outside (red box). The input sequence is passed through an encoder which produces token level embeddings. The CLS token embedding is passed to the WCED decoder, producing two sets of logits the size of the gene vocabulary. The expression levels of all genes are then used to calculate the binary and mean squared error losses.
  \textbf{C.} The package provides a simple scanpy-style API for zero-shot inference on single-cell RNA-seq data. The \texttt{bmfm.inference()} function generates cell embeddings and predictions directly from pre-trained models held on HuggingFace. The results are then stored in the AnnData object allowing users to integrate with their existing workflows.
  \textbf{D.} CellXGene validation corpus wide zero-shot cell type annotation.
  \textbf{E.} Roll-up hierarchically aware zero-shot cell type annotation.}
  \label{fig:pretraining-overview}
\end{figure*}

To address this, we introduce whole-cell expression decoding (WCED), an autoencoder-style pretraining objective in which a shallow decoder must reconstruct the expression state of the entire gene vocabulary from the single [CLS] token embedding produced by the encoder (Figure~\ref{fig:pretraining-overview}A-B).
This bottleneck forces the [CLS] embedding to encode the complete cellular state rather than distributing information across per-gene token representations, connecting to the compression principle underlying variational approaches such as scVI~\cite{lopez2018deep}.
Because the decoder reconstructs the full vocabulary while the encoder sees only a subset, WCED also sidesteps the long-standing zeros problem in TFM design.
With a median of only ${\sim}$1,700 non-zero genes out of a vocabulary exceeding 60,000 in the \cellxgene{} corpus, zeros dominate every expression profile, yet they carry biological signal: a gene that is consistently silenced in a cell type is informative about that cell's identity.
Most TFMs must either discard zeros entirely~\cite{theodoris_transfer_2023}, losing this signal, or include them heuristically in the input~\cite{cui_scgpt_2024,hao_scfoundation_2024}.
WCED supervises on the full expression profile, including zeros, without inflating the input sequence, producing maximally informative representations from limited input.
The WCED decoder outputs both binary detection logits (was each gene expressed?) and continuous expression predictions for nonzero values, trained with binary cross-entropy and mean squared error respectively.

Gene-level error tracking across both WCED and MLM, with classification of genes by functional family and expression program (Supplementary Sections~\ref{sec:gene-family-classification} and~\ref{sec:axis2-regime}), reveals that what determines whether a gene's expression level can be quantitatively predicted is not its biochemical function but the degree to which its level is inferable from the rest of the cell's transcriptomic profile.
Genes whose expression levels co-vary with stable cell-type programs are systematically more predictable than sparsity alone would suggest, while genes whose levels are driven by transient or extra-transcriptomic factors such as dissociation stress or cell-cycle progression are not, across functional families.
This boundary is shared across both objectives ($\rho = 0.858$ rank correlation of family-level learnability), suggesting it reflects a fundamental property of the data rather than a modelling artefact.

Using this insight, we further asked whether explicit cell-type supervision could further improve representations.
However, cell-type labels across studies in curated sources such as CellXGene are annotated at inconsistent levels of resolution: an identical cell may be labeled ``CD8-positive, alpha-beta thymocyte'' in one study and ``T~cell'' in another.
Flat cross-entropy over all observed labels treats these as mutually exclusive classes, discarding the hierarchical relationships among them.
We therefore introduce a hierarchical cross-entropy (HCE) loss that exploits Cell Ontology structure~\cite{tan-cell-ontology-arxiv-2025}, predicting leaf-node probabilities and rolling them up to internal nodes so that coarse and fine labels contribute consistently to the same likelihood (see Methods).
Unlike the unnormalized multi-label formulation in scPRINT~\cite{kalfon_scprint_2025}, our approach yields a probabilistically consistent hierarchy that supports both global leaf-argmax and greedy top-down decoding strategies, enabling both zero-shot cell-type annotation across ontological levels and improved representations during multitask pretraining.
Previous work by scCello~\cite{yuan2024cellontology} incorporated ontology structure through contrastive learning; our formulation instead produces an explicit classifier whose outputs can be queried at any depth of the ontology.
An alternative hierarchical loss was recently described by Cultrera di Montesano et al.~\cite{montesano2026hce} for supervised annotation.
The \bmfmrna{} framework provides a simple API for zero-shot inference (Figure~\ref{fig:pretraining-overview}C), enabling direct application to new datasets, and supports hierarchically-aware cell type annotation at multiple granularity levels (Figure~\ref{fig:pretraining-overview}D-E).

By combining WCED and HCE pretraining, we produce cell-level embeddings that achieve performance comparable to or exceeding state-of-the-art approaches, including scGPT~\cite{cui_scgpt_2024}, Geneformer~\cite{theodoris_transfer_2023}, TranscriptFormer~\cite{pearce_cross-species_2025} and UCE~\cite{rosen_uce_2024}, on zero-shot batch integration and downstream tasks such as cell-type annotation, despite using a smaller model, a more compact representation, and less training data.

We first show that WCED produces superior cell representations despite higher reconstruction error than MLM, then evaluate the combined multitask objectives on zero-shot batch integration across ten datasets, zero-shot cell-type annotation using Cell Ontology structure, and fine-tuned classification on held-out batches.
The methods are implemented in \bmfmrna{}, an open-source modular framework for TFM development, which is described in detail in the supplementary material and accessible at \url{https://github.com/BiomedSciAI/biomed-multi-omic}.

\section{Results}

\subsection{MLM achieves lower pretraining loss, but produces inferior representations}
\label{ref:mlmvswced}

To assess the extent to which the WCED objective provides superior cell representations compared to the standard MLM objective, we trained two Llama models with 12 layers and a hidden size of 384, 37M parameters, trained on 1\% of the \cellxgene data for 10 epochs.
To isolate the effect of the pretraining objective, we first trained matched models on a 1\% subsample; the full multitask models evaluated later use 10\%.
We found that in terms of reconstruction error, MLM consistently obtained lower rescaled MAE than WCED, with rescaled MAE converging to $\sim0.65$ for MLM and $\sim0.90$ for WCED. 

However, this superior reconstruction performance does not translate into improved downstream tasks.
Examining the cell representation quality through Average bio, average batch and F1 score of a logistic regression model trained to predict the cell type shows that in every case the WCED trained model had superior performance, with 0.128 higher F1, 0.149 higher average bio and 0.064 higher average batch (detailed comparison across all datasets in Supplementary Figure~\ref{fig:wced_mlm_zero_shot} and Table~\ref{tab:zero_shot_eval_full}; dataset details in Supplementary Table~\ref{tab:datasets}).

We hypothesize that this surprising mismatch in pretraining and downstream task performance lies in the task difference.
While masked gene prediction can succeed by learning pairwise gene correlations, utilizing thousands of potentially redundant data points to predict the masked value, the WCED bottleneck forces the model to compress the cell state into a transferable representation.

\subsection{Gene-level errors reveal a hierarchy of learnability}

To better understand what exactly the models are learning we track gene-level errors and classify genes along two orthogonal axes: functional family (what the gene does) and expression program (how the gene is regulated), both defined a priori from external databases before examining any model output (Supplementary Sections~\ref{sec:gene-family-classification} and~\ref{sec:axis2-regime}).
The training objective is composed of two distinct tasks: ``detection'' (was the gene expressed in the sample) and ``quantification'' (what was the log-normalized read value of the gene, conditional on it being expressed).

For both WCED and MLM, fewer than 2\% of genes achieve quantification error below a null baseline (predicting the validation mean in all cases).
This difficulty has been observed recently in the context of the Arc Virtual Cell Challenge, where no models significantly outperformed the null MAE~\cite{vcc,divaeinia-vcc2026}.
The detection task performs substantially better, with median family-level ROC-AUC of 0.86 (WCED) and 0.88 (MLM) and no gene family falling below 0.82.
This asymmetry is consistent with findings that binary cell representations capture much of the structure in scRNA-seq data~\cite{bouland-nargb2021} and that shuffling expression values among expressed genes preserves cell-type embedding structure~\cite{moriel-neurips2025}: the learnable signal resides in \emph{which} genes are on, not in their precise levels.
Despite MLM encountering each gene approximately 30-fold less frequently than WCED due to masking, its per-gene rescaled MAE is comparable, suggesting that observation frequency is not the binding constraint on gene-level predictability.

\begin{figure*}[!htb]
    \centering
    \includegraphics[width=\textwidth]{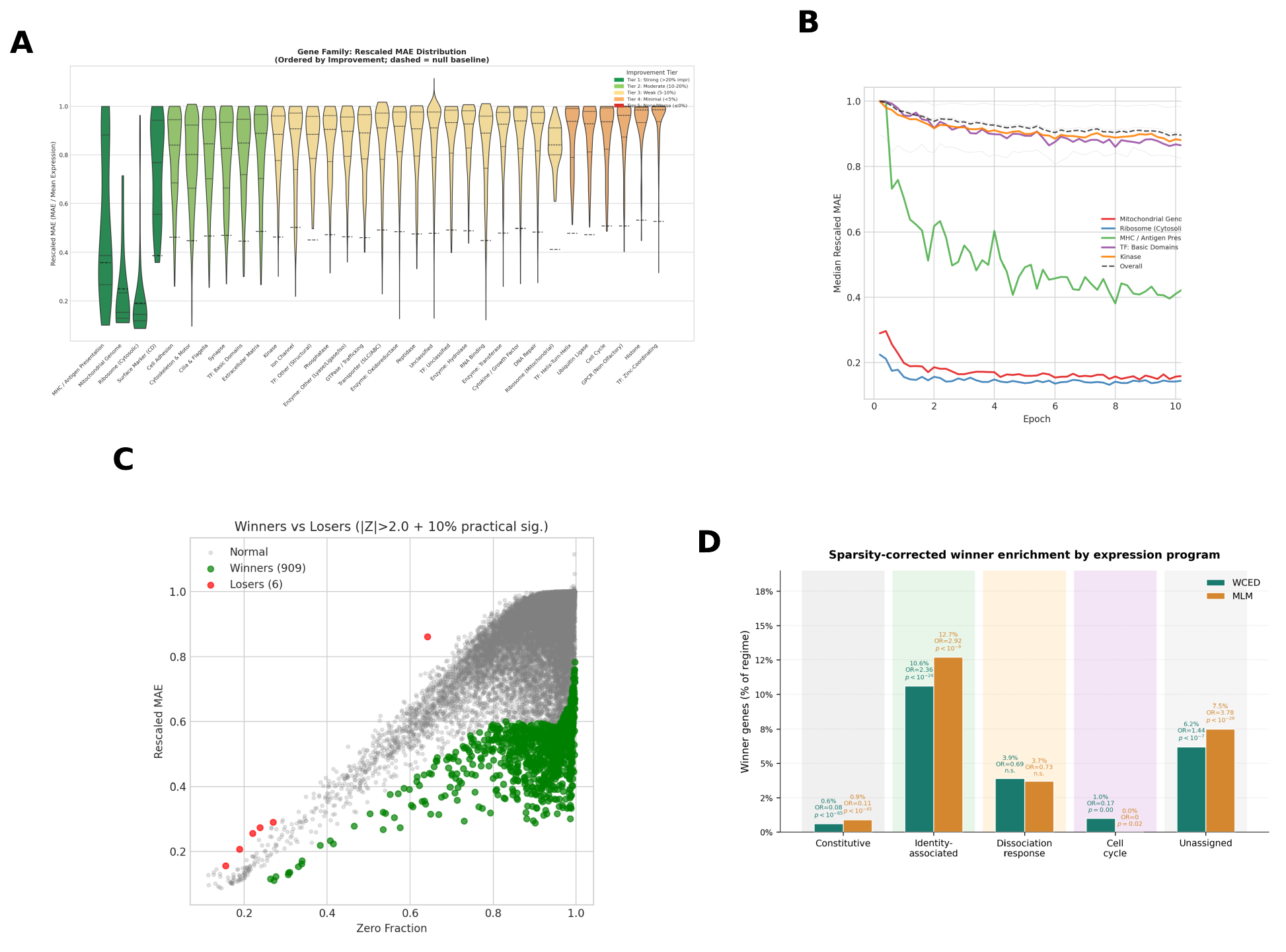}
    \caption{\textbf{Gene-level error analysis reveals a hierarchy of learnability governed by contextual predictability.}
    \textbf{A.}~Distribution of rescaled MAE across 33 gene families, ordered by median improvement over a null baseline (predicting the validation mean). Families cluster into four tiers: Tier~1 (green; mitochondrial-encoded, cytosolic ribosomal, MHC) achieves 20--45\% improvement; Tier~4 (red; zinc-finger TFs, histones) achieves $<$5\%. Violin width reflects gene count. Gene family definitions in Supplementary Table~\ref{tab:gene_families}.
    \textbf{B.}~Per-family training dynamics showing median rescaled MAE over 10 epochs. Constitutive families (mitochondrial genome, cytosolic ribosomes) converge within the first epoch, while most families plateau by epoch~2, indicating that the learnability hierarchy reflects data structure rather than optimization dynamics.
    \textbf{C.}~Winner and loser genes identified by heteroscedastic outlier detection (Supplementary Section~\ref{sec:supp-winners}). After fitting an isotonic regression of rescaled MAE on zero fraction with LOWESS-estimated local variance, genes with $|Z| > 2$ and $>$10\% practical significance are flagged as winners (green; $n = 909$) or losers (red; $n = 6$). Classification is performed across the full sparsity range, but biological interpretation focuses on the sparse regime (zero fraction $> 0.5$), where functionally diverse local comparison groups make Z-scores most meaningful. Winners overwhelmingly cluster in this regime, corresponding to cell-type markers whose expression is contextually predictable despite extreme sparsity. The few losers all fall in the dense regime where interpretability is limited (see Supplementary Section~\ref{sec:supp-winners}).
    \textbf{D.}~Winner enrichment by expression program (Supplementary Section~\ref{sec:axis2-regime}), showing the percentage of genes in each program classified as winners for both WCED and MLM. Identity-associated genes are strongly enriched (WCED: OR $= 2.36$, $p = 7.1 \times 10^{-24}$; MLM: OR $= 2.92$, $p = 5.3 \times 10^{-8}$), while cell-cycle genes (OR $= 0.17$) and dissociation-response genes (OR $= 0.69$, n.s.) are not, confirming that contextual predictability, not biological function per se, determines relative learnability. Note that the low constitutive winner rate reflects the dense regime where Z-score comparisons are among a narrow, homogeneous gene population (see Supplementary Section~\ref{sec:supp-winners}).}
    \label{fig:gene-family-learnability}
\end{figure*}

Two complementary measures of learnability emerge from this analysis (Figure~\ref{fig:gene-family-learnability}):
\emph{Absolute learnability} (raw rescaled MAE) is dominated by expression frequency, and constitutive genes such as mitochondrial-encoded and cytosolic ribosomal genes, achieve the lowest error and the highest rates of beating the null baseline (Figure~\ref{fig:gene-family-learnability}A).
\emph{Relative learnability}, measured by sparsity-corrected Z-scores (Supplementary Section~\ref{sec:supp-winners}), asks whether a gene is predicted better or worse than other genes at the same expression frequency.
This distinction is critical: CD surface markers, for example, show the highest relative winner rate in both models (25.9\% WCED, 20.7\% MLM) yet 0\% beat the null baseline outright, because they are extremely sparse (median zero fraction 0.94) but their expression levels, when nonzero, co-vary tightly with broader immune cell-type programs that the model can read from other genes in the profile.
Training dynamics (Figure~\ref{fig:gene-family-learnability}B) show that constitutive families converge within the first epoch while most gene families hardly reduce the rescaled MAE at all. A notable exception is the MHC presenting family which shows persistent improvement throughout the training.

To test what drives relative learnability, we classified genes into five expression programs: identity-associated, constitutive, cell-cycle, dissociation-response, and unassigned (Supp. Sec. \ref{sec:gene-family-classification}).
Winner genes identified by heteroscedastic outlier detection (Figure~\ref{fig:gene-family-learnability}C) overwhelmingly cluster in the sparse regime (zero fraction $> 0.5$), corresponding to cell-type markers whose expression is contextually predictable despite extreme sparsity.
Identity-associated genes show significant excess learnability (10.6\% winner rate in WCED, OR = 2.36, $p = 7.1 \times 10^{-24}$; 12.7\% in MLM, OR = 2.92, $p = 5.3 \times 10^{-8}$; Figure~\ref{fig:gene-family-learnability}D), while cell-cycle genes (OR = 0.17, $p = 1.6 \times 10^{-3}$) and dissociation-response genes (OR = 0.69, n.s.) are not enriched among winners.
The unifying principle is contextual predictability: genes whose expression levels are inferable from the surrounding transcriptomic context are quantitatively predictable beyond sparsity expectations, while genes whose levels are governed by factors not captured in a static expression snapshot are not.
This hierarchy is consistent across both WCED and MLM (rank correlation $\rho = 0.858$), indicating that it reflects a property of the biology rather than of the architecture.

Expression sparsity (fraction of times the gene was expressed at all) explains 42\% of variance in WCED prediction error ($R^2 = 0.60$ for MLM). Genes expressed in fewer than 20\% of samples are generally not learnable to quantify.
However, substantial variation remains after controlling for sparsity.
Using heteroscedastic outlier detection (Supplementary Section~\ref{sec:supp-winners}), we identified 909 winner genes in WCED ($Z < -2$, $>$10\% practical significance) and 371 in MLM, the vast majority in the sparse regime (zero fraction $> 0.5$) where local comparison groups are functionally diverse and Z-scores are most interpretable.
Of the MLM winners, 76.3\% are also WCED winners, confirming that winner status reflects an intrinsic biological property rather than a model-specific artefact.
Very few genes qualify as losers (6 WCED, 13 MLM; Supplementary Table~\ref{tab:losers}), and all fall in the dense regime (zero fraction $< 0.65$) where the local comparison group is small and homogeneous, precluding confident biological interpretation (Supplementary Section~\ref{sec:supp-winners}).
The highest winner rates belong to families whose members tend to be expressed in narrow, stereotyped cell populations, such as surface markers (CD), synapse, extracellular matrix, cilia \& flagella, and cell adhesion, where their expression levels are tightly coupled to broader transcriptional programs that the model can leverage for prediction.

Dissociation-response genes provide a critical test case. They form a well-characterized, co-activated transcriptional program and are easily detected, but because their expression levels are driven by dissociation severity rather than cell identity, the model cannot predict how much they are expressed from transcriptomic context alone.
This is confirmed by the low rate of winners (OR = 0.69, n.s.) and significantly worse Z-scores than the unassigned baseline ($p = 2.3 \times 10^{-3}$, WCED), consistent with their levels being driven by dissociation protocol and cell vulnerability~\cite{vandenbrink2017single}, factors that leave no systematic trace in the rest of the expression profile.
While the enrichment of cell-type markers among winner genes is broadly consistent with a simple type-versus-state framing, the more precise characterization is that a gene's expression level is learnable when it is redundant with the surrounding transcriptomic context, of which cell-type identity programs are a major cause.
Given that cell-type identity is the dominant learnable signal, we next asked whether providing explicit cell-type supervision during pretraining could further improve representations.

\subsection{Incorporating ontologically aware cell type supervision during pretraining}

Given that cell-type identity is the dominant learnable signal (Section~\ref{ref:mlmvswced}), we combined our hierarchical cross-entropy loss with the reconstructive objectives during multitask pretraining, improving the quality of the representations and enabling zero-shot prediction of Cell Ontology labels at any level of granularity.

\begin{figure}[!htb]
    \centering
    \includegraphics[width=\linewidth]{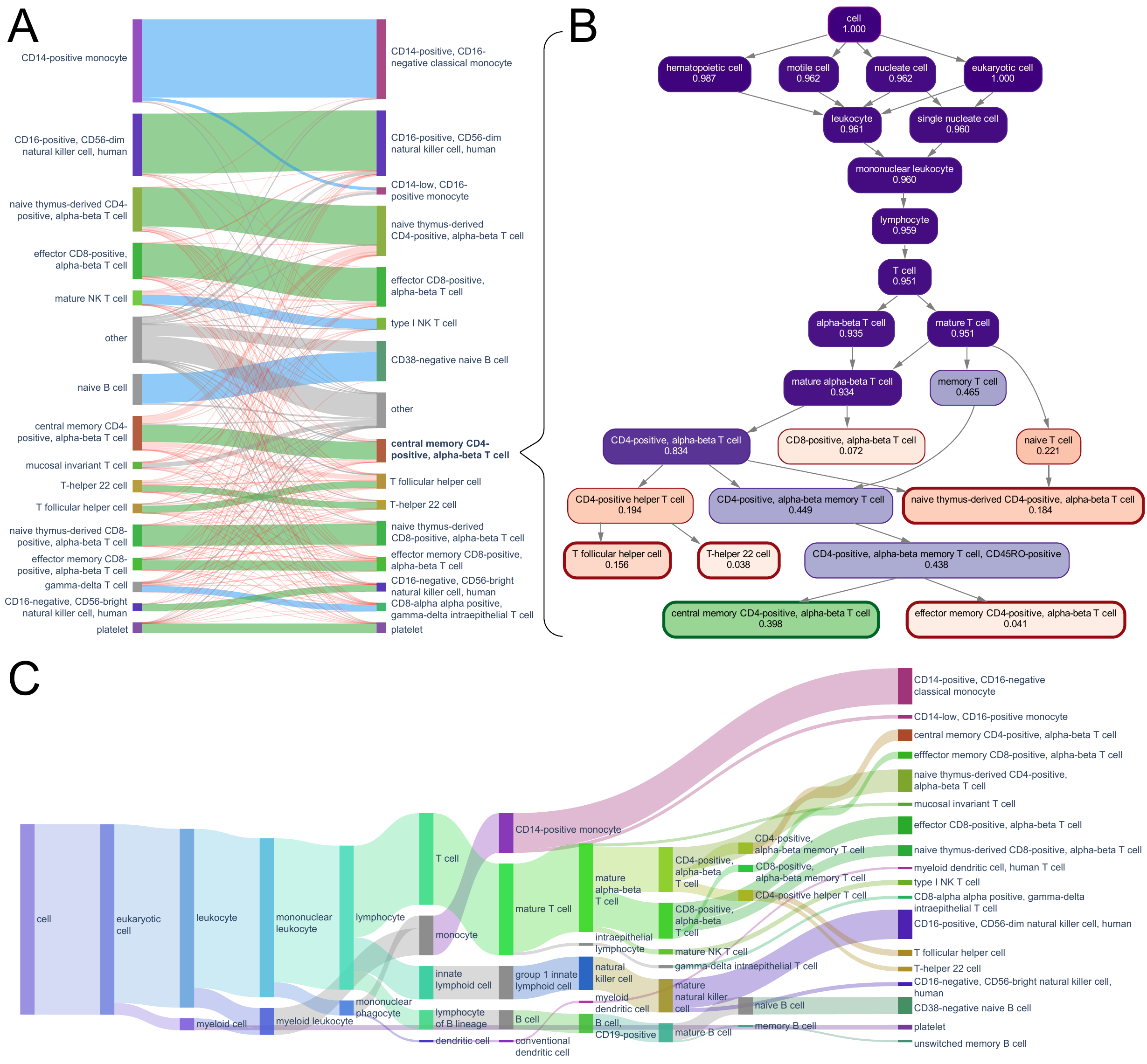}
    \caption{Zero-shot predictions of Cell Ontology labels for Immune Atlas dataset. A. Sankey diagram showing 15 most frequent labels in the test data (left) mapped to 15 most frequent zero-shot labels from the model (right). 'Other' pools all other labels in the test dataset. Flow colors show exact label matches (green), descendant label matches (blue), label misses (red), and other (grey). B. Graph of nodes in the Cell Ontology with mean probability $>0.035$ for cells with ground truth label ``central memory CD4-positive, alpha-beta T cell'' (2327 cells). The model predicts leaf node (bold outlines) probabilities directly, and non-leaf node probabilites are the sum of all leaf node descendants. Darker colors show high probabilities, colors show exact match (green), ancestor match (purple), misses (red). C. Flow of labels assigned to cells, starting from the ontology root (`cell', left) and iteratively selecting the highest probability child for each cell until a leaf (right side) node is reached.}
    \label{fig:zeroshot_ontology}
\end{figure}

On the \texttt{\detokenize{immune_atlas}} dataset, which uses the same cell-type vocabulary as the training corpus, predicted cell types were frequently either exact matches to the ground truth or more fine-grained descendant labels (green and blue flows in Figure~\ref{fig:zeroshot_ontology}A).
Because leaf probabilities roll up to ancestor nodes, the model's predictions can be examined at any level of ontological coarseness.
For example, for cells with the ground-truth label ``central memory CD4-positive, alpha-beta T cell'' the correct label is the most probable leaf (mean probability 0.398), but substantial probability mass is distributed across nearby leaves such as ``na\"ive thymus-derived CD4-positive, alpha-beta T cell'' and ``T follicular helper cell'' (Figure~\ref{fig:zeroshot_ontology}B).
Among ancestor nodes, confidence increases more sharply for ``CD4-positive, alpha-beta T cell'' than for ``memory T cell,'' indicating that the model resolves the CD4 lineage confidently while remaining uncertain about the memory-vs-naive distinction within it.
This roll-up behaviour provides a natural mechanism for identifying the level of granularity at which cell-type assignments become reliable, and may reflect genuine heterogeneity in the labeled population rather than model error alone.
The hierarchical decoding process (Figure~\ref{fig:zeroshot_ontology}C) iteratively selects the highest probability child at each level, starting from the ontology root until reaching a leaf node.

Having established that both WCED and the hierarchical cell-type objective independently improve representations, we next evaluated their combination on zero-shot batch integration.

\subsection{Comparing contribution of WCED and supervised cell-type training}\label{sec:1pct-multitask}
To assess the relative advantages of WCED and hierarchical cross entropy supervision for cell types during pre-training, we used the same architecture and trained with a multitask objective for the same number of epochs.
We find that the introduction of the supervised hierarchical cell-type training mostly closes the gap between the WCED and MLM models, producing a small boost for WCED and a very large boost for MLM.
This is consistent with our hypothesis that the WCED objective is producing effective cell type representations.
Without HCE, WCED produces markedly better representations than MLM (F1 0.877 vs.\ 0.742, $d = 1.22$, $p < 0.003$, $n = 12$ shared datasets), consistent with the information bottleneck hypothesis: the CLS decoding constraint forces compression of all inputs that masked prediction does not require.
Adding HCE eliminates this gap, with MLM gaining $+0.129$ F1 ($d = 1.14$, $p < 0.002$) and $+0.222$ Avg Bio ($d = 2.61$, $p < 5 \times 10^{-4}$), bringing it to parity with WCED on all metrics (all $p > 0.1$).
WCED F1 is unchanged ($-0.011$, n.s.) but batch integration improves (Avg Batch $+0.083$, $d = 1.23$, $p < 5 \times 10^{-4}$).
The cell-type loss thus provides, through explicit supervision, the cell-type encoding that WCED achieves architecturally.
HCE is thus an excellent task, but limited to datasets where such labels are available and privileging the "textbook" cell-types only.
MLM and CLS alone are ill-prepared for biological questions beyond standard cell-types, such as recognizing the effect of drug perturbations on transcriptomic profiles.

On the basis of these results, we trained both WCED MULTITASK and MLM MULTITASK on 10\% of \cellxgene{} (${\sim}$6.2M cells) as detailed in Sec \ref{checkpoints}.
These are the checkpoints evaluated in all subsequent analyses and released with the \bmfmrna{} framework.
At this scale, scaling from 1\% to 10\% yielded a small additional F1 gain ($+0.008$--$0.012$, $p < 0.025$) with no significant change in integration metrics (Avg Bio, Avg Batch: all $p > 0.05$), indicating that the representation quality is affected more by training objective than by corpus size.

\subsection{Zero-shot cell representation performance}

\begin{figure}[!htb]
  \centering
  \includegraphics[width=\textwidth]{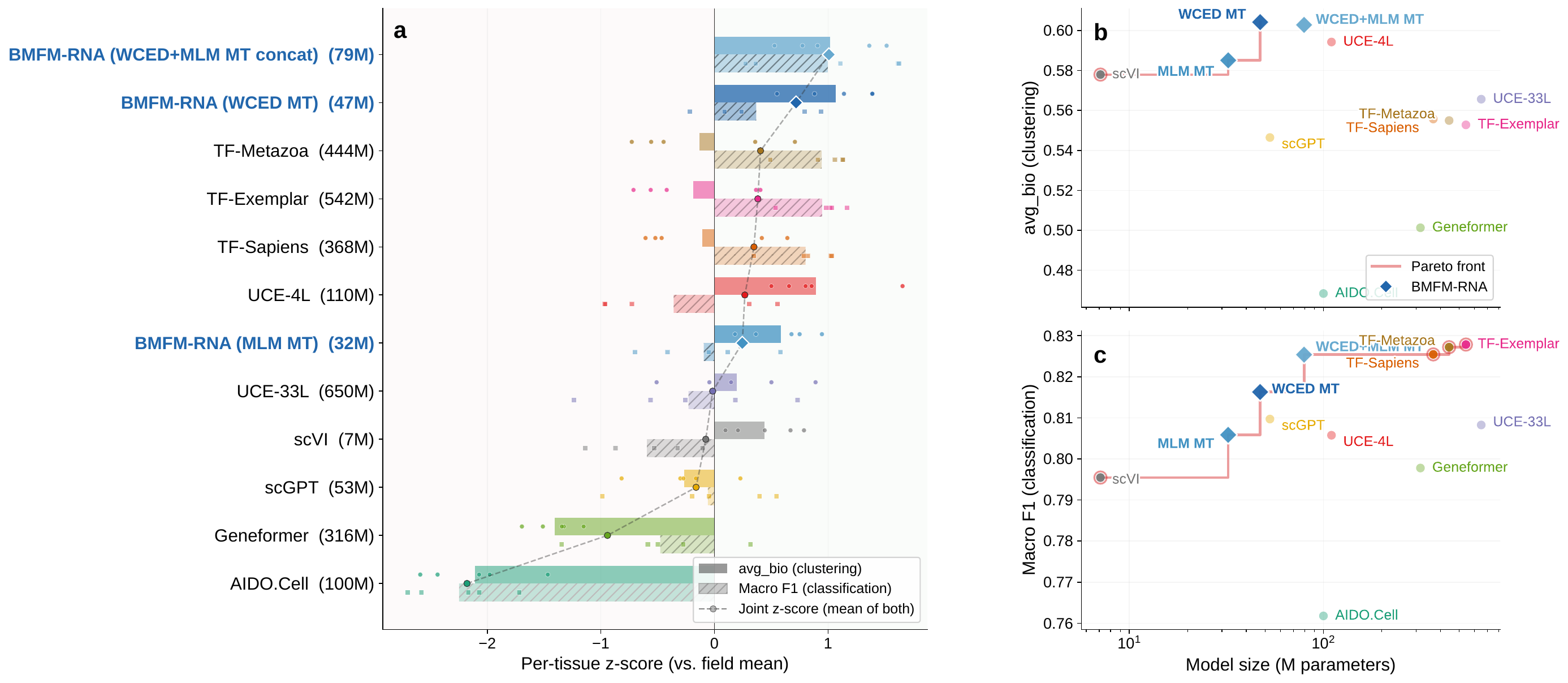}
\caption{\textbf{Cross-benchmark evaluation of cell embedding models on CZ CELLxGENE clustering and classification tasks.}
\textbf{(a)} Per-tissue $z$-scores relative to the field mean for unsupervised clustering quality (Avg Bio = mean of ARI, NMI, and silhouette score; solid bars) and supervised cell type classification (Macro F1; hatched bars) across five Tabula Sapiens v2 tissues~\cite{tabula_sapiens_v2}. All results were obtained using the CZ CELLxGENE benchmarking suite~\cite{cz-benchmarks} with published data splits and evaluation protocols. Individual dots show per-tissue $z$-scores; diamond and circle markers connected by a dashed line show the joint $z$-score (mean of Avg Bio and Macro F1), which determines the vertical ordering. \bmfmrna{} multitask checkpoints (32--79M parameters) occupy the top ranks, with the WCED+MLM multitask concatenation achieving the highest joint score. TranscriptFormer variants~\cite{pearce_cross-species_2025} (368--542M parameters) rank highly on classification but fall below the field mean on clustering, consistent with a 2048-dimensional representation that preserves input signal without reorganizing it into biologically structured geometry. \textbf{(b, c)} Performance versus model size for Avg Bio (b) and Macro F1 (c). The Pareto front (coral line and rings) traces the best performance achievable at each parameter budget. \bmfmrna{} checkpoints define the Pareto front for Avg Bio above scVI~\cite{lopez2018deep}, achieving the highest clustering scores in the comparison at 6--17x fewer parameters than TranscriptFormer. On Macro F1, \bmfmrna{} (WCED+MLM multitask concatenation) reaches the Pareto front at 79M parameters, matching TranscriptFormer's classification accuracy at roughly one-fifth the model size.}
  \label{fig:czi_zscore_benchmark}
\end{figure}

We bench\-marked BMFM-RNA against nine published models using the cz-benchmarks~\cite{cz-benchmarks} evaluation suite on identical Tabula Sapiens 2.0 data and splits, ensuring no advantage from custom evaluation choices.
Given that the WCED MULTITASK and MLM MULTITASK models are trained to optimize slightly different objectives, we also tested the representations produced by concatenating the embeddings produced by both models, creating a 768 dimensional embedding referred to as BMFM CONCAT in the figures.
The results (Figure~\ref{fig:czi_zscore_benchmark}a-c and Supplementary Tables~\ref{tab:czi_f1}, \ref{tab:czi_accuracy}, \ref{tab:czi_cell_clustering_ari}, \ref{tab:czi_cell_clustering_nmi}, \ref{tab:czi_cell_clustering_ss}) expose a tension between two standard evaluation axes.
TranscriptFormer leads on supervised cell type classification (Macro F1 up to 0.828) but falls to rank 7-9 on average bio, the unsupervised clustering composite. BMFM-RNA shows the inverse: it leads on average bio while trailing TranscriptFormer on F1 by only 0.003.
TranscriptFormer's 2048-dimensional embeddings are large enough to preserve enough of the input expression profile to support accurate downstream probing without necessarily learning to compress it into biologically meaningful structure.
The average bio metrics, which evaluate intrinsic embedding geometry rather than information retention, expose this limitation.
BMFM-RNA achieves comparable classification from 384-dimensional representations, one-fifth the size, while producing the field's best-structured embedding space.
The concat embeddings produce superior results to the individual models, at times surpassing the far larger TranscriptFormer embeddings while preserving average bio performance better than any non-BMFM model on the benchmark as of March 2026.
This is the profile expected of a model that has learned genuine abstraction of transcriptomic state rather than high-fidelity storage of its input.
The same trend holds when we consider the relationship between model size and zero shot macro F1 (Figure~\ref{fig:czi_zscore_benchmark}b-c), finding that BMFM-RNA models consistently outperform other models in their size class at the supervised task and define the Pareto front for clustering quality.

We further explore the zero-shot performance by exploring the more diverse set of datasets collected by  \scEval{}~\cite{liu_evaluating_2024} comparing our performance to \scGPT{} using the weighted average score of both the batch removal score (AvgBatch) and the bio-conservation score (AvgBio) to balance biological relevance and batch consistency. For more details on the metrics, we refer to \cite{cui_scgpt_2024,yuan2024cellontology}. 
Supplementary tables \ref{tab:Weighted_BIO_Batch}, \ref{tab:Average_Bio} and \ref{tab:Average_Batch} summarize zero-shot AvgBio, AvgBatch and weighted average scores of the two multitask models, the concatenated embeddings and \scGPT{} on 10 datasets from \scEval{}'s batch integration task (dataset details in Supplementary Table~\ref{tab:datasets}). 
We find that both BMFM-RNA multitask checkpoints outperform scGPT on average (MLM MULTITASK: 0.747, WCED MULTITASK: 0.748 vs. scGPT: 0.730), despite being trained on approximately 10x less data with a smaller model.
We find that the addition of the cell-type ontologically aware hierarchical cross entropy task alongside the reconstructive task compensates for the poor quality of simple MLM training discussed above, producing a checkpoint competitive with WCED.
Given that batch correction is uniformly strong across all models (AvgBatch 0.87--0.96, Supplementary Table~\ref{tab:Average_Batch}), the performance gap is driven almost entirely by bio-conservation (Supplementary Table~\ref{tab:Average_Bio}), further evidence that BMFM-RNA embeddings preserve biologically meaningful structure across diverse datasets.
No single multitask checkpoint dominates uniformly, but BMFM-RNA multitask models account for the majority of per-dataset wins.

\subsection{Fine-tuned Cell Type Annotation}
\begin{figure}[!htb]
    \centering
    \includegraphics[width=\linewidth]{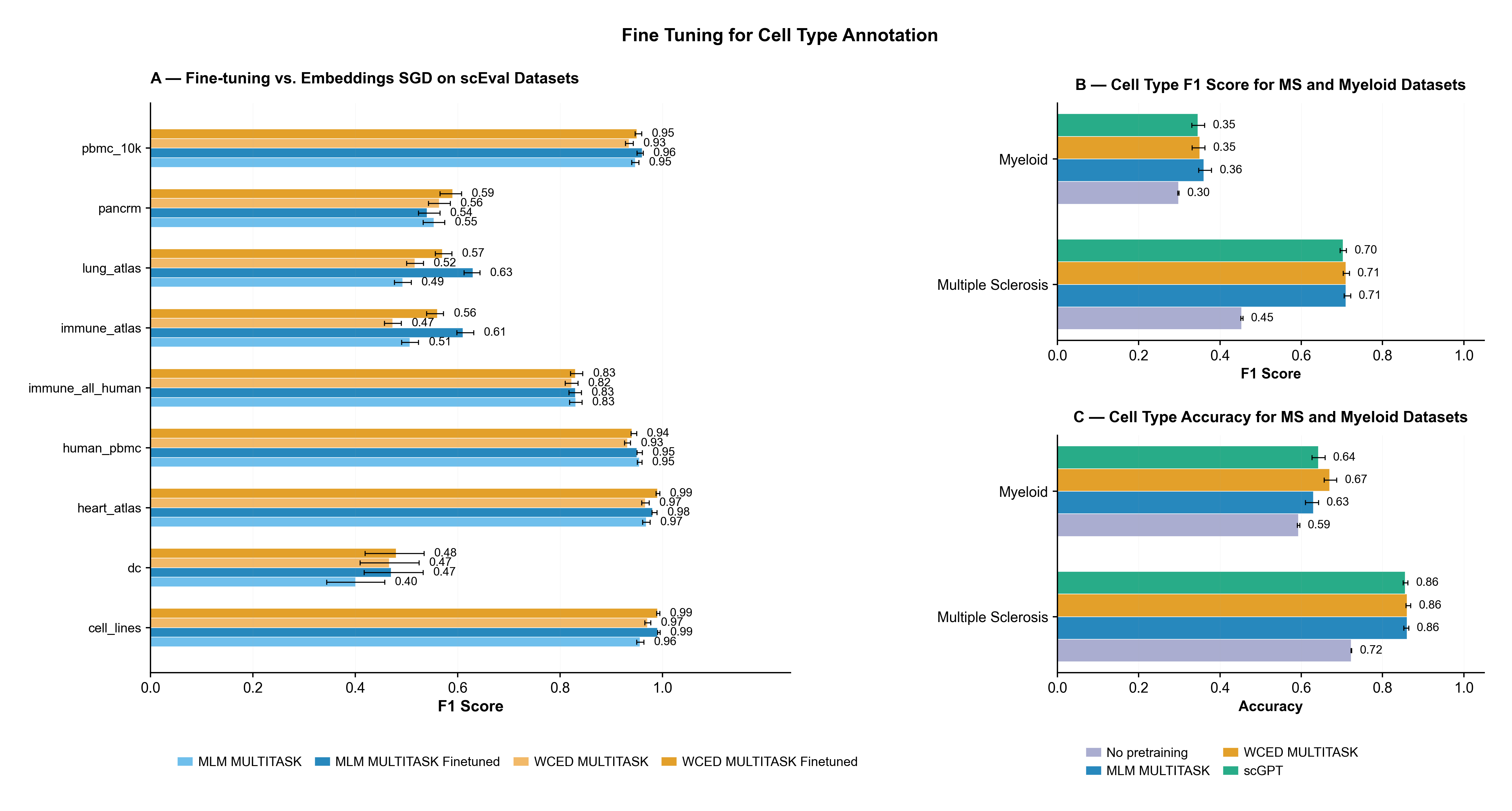}
    \caption{\textbf{Cell Type Annotation Fine Tuning results.}
    \textbf{A.} F1 scores of classifiers for 9 scEval datasets split by batch with binomial confidence interval at 95\%. For each dataset we fine tune each of the models over the Cell Type Annotation task with unfrozen encoder for five epochs and compare to an SGD classifier trained over extracted model embeddings. The fine-tuned models achieve better performance with average F1 improvement of 0.028 for MLM MULTITASK and 0.039 for WCED MULTITASK.
    \textbf{B.} Classification F1 score with binomial confidence interval at 95\% for fine tuning the Myeloid and Multiple Sclerosis datasets reported by scGPT using \bmfmrna{} models.
    \textbf{C.} Classification accuracy with binomial confidence interval at 95\% for fine tuning the Myeloid and Multiple Sclerosis datasets reported by scGPT using \bmfmrna{} models. Results were calculated on the same splits as scGPT used, based on the files shared in their paper.}
    \label{fig:combined_ft}
\end{figure}

While zero-shot performance demonstrates representation quality, fine-tuning evaluates adaptability to specific datasets, and can produce label predictions of greater quality than zero-shot by leveraging the full expressivity of the transformer architecture.
We finetuned the models for 9 of the datasets introduced in \scEval{}, with evaluations calculated on held out batches (Supplementary Table~\ref{tab:datasets}).
We also trained a linear classifier over zero-shot embeddings as a baseline. In addition we finetuned the models for the two datasets, namely Multiple Sclerosis (MS) and Myeloid introduced in \scGPT{} (Supplementary Table~\ref{tab:datasets}). For these two datasets, we use the processed files with the same splits, as shared by \scGPT. Thus, the MS is an out-of-distribution testing scenario, and the Myeloid testing set includes cell types that are missing in the training process.
Classification performance on \scEval{} datasets is measured with F-1 score and reported in Figure~\ref{fig:combined_ft}A.

We find that fine-tuning consistently improves over frozen-embedding classifiers for both multitask checkpoints, with average F1 gains of $+0.028-0.039$.
Gains are largest on datasets with moderate baseline performance (\texttt{lung\_atlas}, \texttt{immune\_atlas}, \texttt{pancrm}), while near-ceiling datasets (\texttt{cell\_lines}, \texttt{heart\_atlas}) leave little headroom.
Most published foundation models do not provide reproducible fine-tuning results, because the same dataset can have very different results due to preprocessing or split definition.
One of the few exceptions was scGPT which shared the exact data used for the Myeloid and Multiple Sclerosis datasets.
In Figure~\ref{fig:combined_ft}B-C we compare with scGPT published results and find both MLM MULTITASK and WCED MULTITASK, trained for four epochs with unfrozen encoder, matching or exceeding scGPT within confidence intervals.

\subsection{Prediction of Drug Mechanisms of Action Using Foundation Model Embeddings}

\begin{figure}[!htb]
    \centering
    \includegraphics[width=\linewidth]{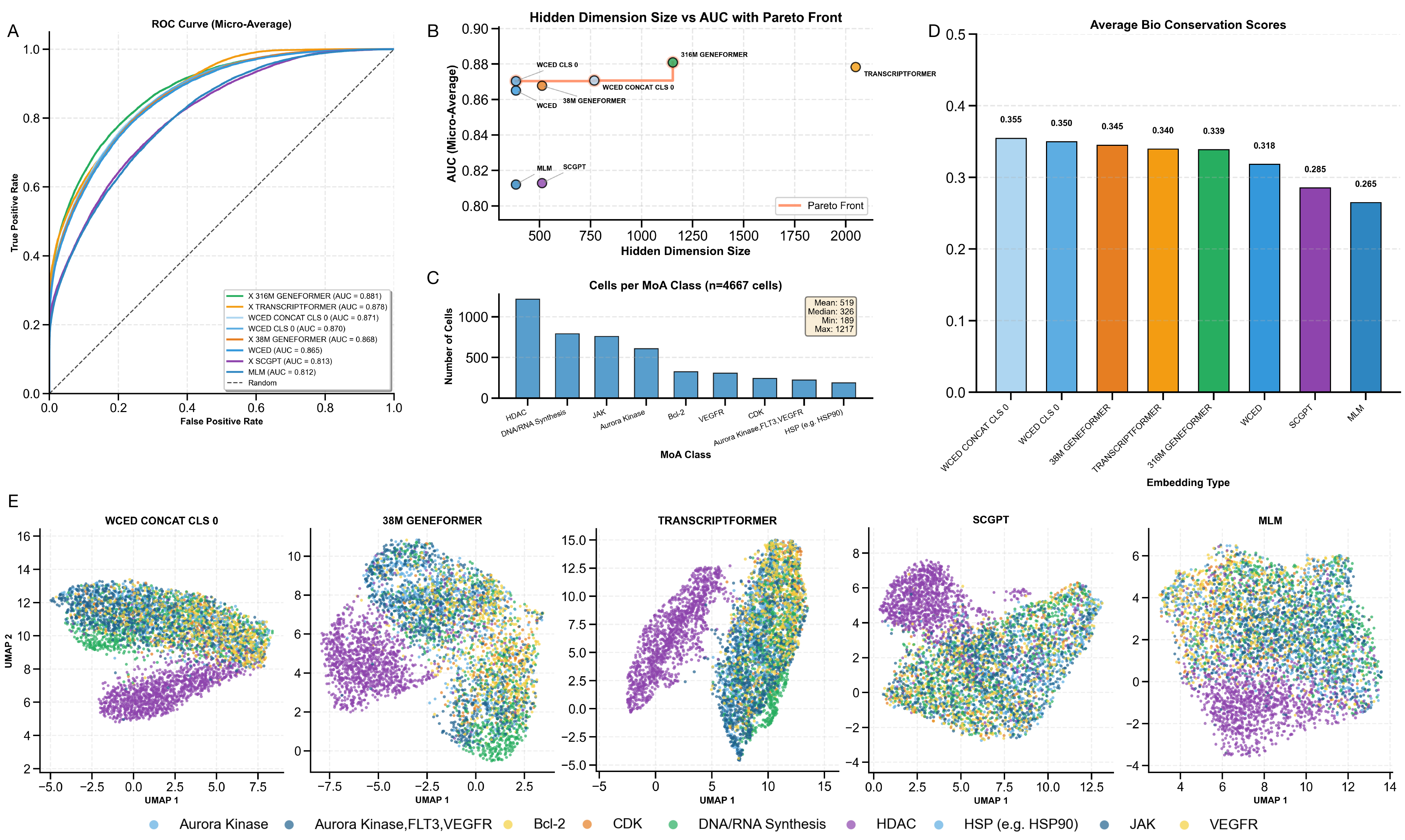}
    \caption{\textbf{Sciplex3 mechanisms of action prediction task results.}
      \textbf{A.} Predictive performance ROC-Curve of a linear probing classifier (Logistic Regression) for the MoA task with stratified drug group cross-validation.
      \textbf{B.} Hidden size vs AUC with Pareto front shown by the coral staircase.
      \textbf{C.} Cells per MoA class from the Sciplex3 datasets. In total there are 9 MoA with 37 unique drugs.
      \textbf{D.} Average Bio scores for each models vectors (BMFM models represented by in blue).
      \textbf{E.} UMAP presentations of the top embeddings per model type with labeled MoA classes.
    }
    \label{fig:sciplex_results_fig}
\end{figure}

Most assessments of TFM representation quality focus on the ability of the representations to recapitulate known cell types.
However, there is significant biologically meaningful transcriptomic variation that is present beyond the cell type distinction: disease status, gene perturbation response and drug perturbation response are important examples.
While supervised pretraining with cell ontology labels would naturally lead to improved representations with respect to ``textbook'' cell type classes, it may not be effective for other biologically meaningful tasks.
To evaluate this questions, we performed zero-shot mechanism of action (MoA) classification on the Sci-Plex~\cite{srivatsan2020sciplex} drug screening dataset (Figure~\ref{fig:sciplex_results_fig}). We compared our models against our baseline MLM-MT model in addition to several state-of-the-art single-cell foundation models, including Geneformer (38M and 316M parameter variants), scGPT and Transcriptformer. The task required predicting the mechanism of action for drug-treated A549 cells based solely on their transcriptional profiles, using stratified drug group cross-validation to ensure generalization to unseen compounds within known MoA classes (Figure~\ref{fig:sciplex_results_fig}C). The performance of the models was evaluated on both their ability to predict the MoA class from the vectors and the extent to which the learned embeddings recapitulate the MoA classes, using biological conservation metrics.
We expect that drugs of a similar MoA class will cause transcriptomic responses which should be clustered together, even if all of the examples are of the same cell type, with respect to the cell-type ontology.

For the MoA classification task, the 316M parameter Geneformer achieved the highest AUC of 0.881, followed closely by Transcriptformer at 0.878. Our WCED concat CLS variant achieved an AUC of 0.871, representing a 1.0\% difference when compared to Geneformer 316M. In contrast, scGPT achieved 0.813 and our MLM-MT embeddings achieved 0.812. These results demonstrate that WCED models achieve competitive classification performance while using substantially fewer parameters and smaller hidden size than the top-performing models with the other variants WCED CLS variant achieved 0.870 and the base WCED quantification model achieved 0.865 (Figure~\ref{fig:sciplex_results_fig}A). The 1.0-1.6\% point differences in AUC between our best WCED variant and the largest foundation models represent modest trade-offs in task-specific performance. Notably, our WCED models outperform the 38M Geneformer (0.868) and substantially exceed scGPT (0.813), despite scGPT having significantly more parameters. The MLM-based embeddings' lower performance (0.812) highlights the importance of the weighted decoder architecture for capturing drug-induced transcriptional changes.

Using scib's framework we quantified how well each model's embeddings preserve known biological structures. These metrics combine Normalized Mutual Information (NMI cluster/label), Adjusted Rand Index (ARI cluster/label), and Average Silhouette Width (ASW label) into an average biological conservation score (avgBIO). The WCED concat CLS model achieved an avgBIO score of 0.355, followed by WCED CLS at 0.350, 38M Geneformer at 0.345, Transcriptformer at 0.340, 316M Geneformer at 0.339, base WCED at 0.318, scGPT at 0.285 and MLM at 0.265 (Figure~\ref{fig:sciplex_results_fig}D-E). The WCED concat CLS model's 1.6 percentage point advantage over the 316M Geneformer and 1.5 percentage point advantage over Transcriptformer indicates improved preservation of biological structure.

The WCED models use a substantially smaller hidden size of 384, compared to that of the representations used by scGPT (512), Geneformer (38M: 768 and 316M: 1152) and Transcriptformer (2048) (Figure~\ref{fig:sciplex_results_fig}B). Despite this reduced dimensionality, WCED models achieve competitive classification performance (AUC: 0.865-0.871) and the highest biological conservation scores (avgbio: 0.318-0.355). The key finding is that our parameter-efficient WCED architecture achieves competitive biological structure preservation compared to substantially larger models, while maintaining competitive performance on the MoA task.
We hypothesize that the improved performance of WCED over MLM in this case is due to the fact that the cell representation is required to compress maximally informative reconstruction information alongside the cell type ontology.
Thus when we ask questions about representation quality outside the cell-type ontology, the added information of the WCED training produces improved results.
As we noted in the discussion of the linear probe task above, for models with very large hidden sizes, the ability to perform well on a linear probe task may reflect the added capacity of the model to store the raw data, regardless of whether it has learned biologically meaningful embeddings.


\section{Discussion}
In this work we introduce whole-cell expression decoding (WCED), a pretraining objective that forces the model to reconstruct the full gene vocabulary from the cell's [CLS] embedding, and show that it produces superior cell representations compared to masked language modeling despite achieving higher reconstruction error.
We further show that ontology-aware hierarchical cell-type annotation provides a complementary route to similar improvement, and that combining the two yields zero-shot and fine-tuned performance competitive with or exceeding scGPT using a smaller model trained on substantially less data.

Utilizing gene-level error analysis, combining functional family and expression program classifications defined entirely from external databases (Supplementary Sections~\ref{sec:gene-family-classification} and~\ref{sec:axis2-regime}), we find a persistent distinction between what is and is not learnable from static transcriptomic data and plausible biological explanations for it.
Absolute learnability as measured by raw prediction error is dominated by expression frequency: constitutive genes achieve the lowest MAE simply because the model observes many non-zero training examples.
We identify winner genes by sparsity-corrected Z-scores which are enriched in identity-associated expression programs (OR = 2.36, $p = 7.1 \times 10^{-24}$ in WCED) while being depleted among cell-cycle (OR = 0.17, $p = 1.6 \times 10^{-3}$) and dissociation-response programs.
This indicates that a gene's expression level is learnable when it co-varies with broader transcriptional programs visible elsewhere in the cell's profile, making it inferable from transcriptomic context.
Cell-type identity programs are the principal source of such contextual redundancy, which explains both why identity-associated genes are enriched among winners and why the models reliably predict \emph{whether} a gene is expressed (median family-level ROC-AUC 0.86--0.88, where presence/absence is tightly coupled to cell type) but struggle with \emph{how much} it is expressed (where levels reflect transient or extra-transcriptomic factors).
This connects to findings that binary cell representations capture much of the information in scRNA-seq data~\cite{bouland-nargb2021} and that shuffling expression values among expressed genes preserves cell-type embedding structure~\cite{moriel-neurips2025}.
Consistent with this view, BMFM-RNA embeddings readily separate disease-associated astrocytes from homeostatic astrocytes in Alzheimer's disease tissue without additional fine-tuning (Supplementary Figure~\ref{fig:astrocyte}), suggesting that pretrained cell representations can support disease subtype discovery in neurodegeneration cohorts (Supplementary Section~\ref{sec:alzheimers-daa}; dataset details in Supplementary Table~\ref{tab:datasets}).

This boundary explains an apparent contradiction in the TFM literature.
Current models perform well on annotation and batch integration~\cite{cui_scgpt_2024,theodoris_transfer_2023,hao_scfoundation_2024} while failing to outperform simple baselines on gene expression level quantitative prediction~\cite{ahlmann-eltze_deep_2024,kedzierska_assessing_2023,vcc,divaeinia-vcc2026}.
Our work amplifies these recent findings, that practitioners should exercise caution on tasks requiring quantitative
gene-level accuracy, especially gene and drug perturbation response.

The WCED bottleneck does not eliminate this problem but extracts more useful representations from the learnable regime.
By requiring reconstruction of the full vocabulary from a single embedding, it prevents MLM's shortcut of succeeding through local pairwise correlations without compressing a transferable representation.
We see the added value beyond questions of cell type, where WCED cell representations effectively segregate drug perturbed transcriptomic responses better than MLM and other models.
This compression principle connects to variational approaches such as
scVI~\cite{lopez2018deep} or the bottleneck learning explored in UCE~\cite{rosen_uce_2024} or scPRINT~\cite{kalfon_scprint_2025,kalfon_scprint2_2025}.
Whether other strategies from the autoencoder domain would further improve quality is an open question that can be tested via the modular \bmfmrna{} framework.
We find that adding hierarchical cross-entropy loss of cell-types produces MLM-trained models that are competitive with WCED, a finding complemented by concurrent work showing that an alternative hierarchical cell-type loss improves supervised classifier robustness~\cite{montesano2026hce}. 
However, it must be noted that this task is limited to datasets with already labeled cell-types which match the cell type ontology.
WCED, on the other hand, can produce high quality representations across any training data, and can even be applied to other modalities such as bulkRNA and epigenetics like methylation~\cite{ying2024methylgpt} or CHIPseq.

Our findings suggest that scaling current architectures on static scRNA-seq is unlikely to make state-responsive genes learnable at the quantification level, because the determining signals may be absent from the input.
Dissociation-response genes illustrate this sharply: despite being a well-characterized, co-activated transcriptional program, they are not enriched among winner genes.
We hypothesize that this is because their expression levels are driven by dissociation protocol and cell vulnerability rather than by the stable transcriptional programs that make cell-type marker levels contextually predictable.
Modeling these continuously graded processes will likely require data modalities that capture causal regulatory information, such as spatial transcriptomics, time-resolved sampling, or epigenetics.

The gene universe covers only human protein-coding genes.
Preliminary experiments with mouse data mapped via orthologs (Supplementary Table~\ref{tab:mouse_zeroshot_combined_metrics}) suggest that the learned representations transfer across species, and systematic cross-species evaluation is an important further direction.
All evaluations use scRNA-seq; spatial, multi-modal, and perturbation-response benchmarks are important future directions.

The most consequential open question is whether this learnability boundary is absolute or permeable.
That two different objectives reflecting different inductive biases agree on which genes' levels are contextually predictable suggests the boundary stems from the data generating process itself, not from architectural choices.
If the boundary is absolute, it would imply that cell-type identity programs generate expression levels that are redundant with the broader transcriptomic context, while cell-state variation is driven by factors such as spatiotemporal position, stochastic regulatory dynamics, or environmental stimuli that are fundamentally not recoverable from a single expression snapshot.
Testing this directly, by incorporating modalities that capture the regulatory processes upstream of expression levels, is a critical next step.
All configurations, checkpoints, and evaluation protocols are publicly released at \url{https://github.com/BiomedSciAI/biomed-multi-omic} to make this and future investigations reproducible.

\section{Methods}

\subsection{\biomedmultiomic{} software package}
The \biomedmultiomic{} framework with which we implemented \bmfmrna{} is a modular, open-source software engine for training and evaluating transcriptomic foundation models, built on PyTorch Lightning, Hugging Face Transformers, and Hydra.
Its design separates data ingestion, tokenization, model architecture, training objectives, and evaluation into independently configurable components, enabling systematic exploration of TFM design choices through configuration alone.
Full architectural details, including data pipeline design, tokenization, and reproducibility infrastructure, are provided in Supplementary Section~\ref{sec:si-software} and illustrated in Supplementary Figure~\ref{fig:si-package-overview}.
The package can be accessed freely at  \nopagebreak{\url{https://github.com/BiomedSciAI/biomed-multi-omic}}.

\subsection{Hierarchical Classification via Cell Ontology}
\label{sec:hierarchical_loss}
To leverage the structured relationships between cell types defined in the Cell Ontology (\url{https://obofoundry.org/ontology/cl.html}), following prior work \cite{kalfon_scprint_2025, yuan2024cellontology}, we represent the ontology as a directed acyclic graph (DAG) $\mathcal{G}=(V,E)$. Each node $v\in V$ corresponds to a biological cell type, and each directed edge encodes a parent--child relation (e.g., \texttt{is\_a}).

Let $\mathcal{V}$ denote the set of ontology nodes and $\mathcal{L}\subseteq \mathcal{V}$ the set of leaves. The model predicts \emph{only leaf labels}: given leaf logits $\mathbf{z}\in\mathbb{R}^{|\mathcal{L}|}$, we define leaf probabilities
\[
s_i=\mathrm{softmax}(\mathbf{z})_i,\qquad i\in\mathcal{L},
\]
and the predicted label is the most likely leaf,
\[
\hat{\ell}=\arg\max_{i\in\mathcal{L}} s_i.
\]
When evaluation or supervision is provided at an internal node, we compare the predicted leaf $\hat{\ell}$ to that internal node through the ontology relations. In particular, for any node $v\in\mathcal{V}$ we define the probability mass assigned to the induced subtree rooted at $v$ as
\[
p(v)=\sum_{i\in \mathcal{D}(v)\cap \mathcal{L}} s_i,
\]
where $\mathcal{D}(v)$ denotes the descendants of $v$ (including $v$ itself when $v$ is a leaf).

\subsubsection{Hierarchical cross-entropy loss}
We consider a classification task over the $K$ terminal nodes (leaves) of $\mathcal{G}$ and let $\mathbf{z}\in\mathbb{R}^K$ denote the model logits over leaves for a given input. Unlike flat classification, the ground-truth label $l$ may correspond to an internal ontology node rather than a specific leaf.

We define the probability assigned to label $l$ as the total probability mass of its descendant leaves. Let $\mathbf{w}\in\{0,1\}^K$ be a binary mask with $w_i=1$ if leaf $i$ is a descendant of $l$ in $\mathcal{G}$ and $w_i=0$ otherwise. Then
\begin{equation}
p(l)=\frac{\sum_{i=1}^K w_i e^{z_i}}{\sum_{j=1}^K e^{z_j}}.
\end{equation}
The hierarchical cross-entropy (HCE) loss is the negative log-likelihood of this cumulative probability:
\begin{align}
\mathcal{L}_{\text{HCE}}
&= -\log p(l) \\
&= -\log\!\left(\sum_{i=1}^K w_i e^{z_i}\right) + \log\!\left(\sum_{j=1}^K e^{z_j}\right).
\end{align}
The second term is a standard \texttt{LogSumExp}. For the first term, we use a masked (weighted) \texttt{LogSumExp} to ensure numerical stability. If a label is marked as \emph{unknown} or cannot be mapped to the ontology, then $\mathbf{w}=\mathbf{0}$; in this case we set the contribution of the example to zero via an indicator ${\mathbb{1}}_{\|\mathbf{w}\|_0>0}$.

Our formulation differs from \textsc{scPRINT}~\cite{kalfon_scprint_2025} in a key structural aspect. 
\textsc{scPRINT} also employs a leaf-only formulation in which the model produces logits exclusively for the leaf nodes of the ontology. However, each leaf node is associated with a raw logit that is passed through a sigmoid independently, meaning that every leaf is treated as a separate binary prediction. As a consequence, there is no normalization across leaves. When the ground-truth label corresponds to a specific leaf node, the loss encourages the sigmoid output of that leaf to approach $1$, while simultaneously pushing the sigmoid outputs of all other leaves toward $0$. When the ground truth corresponds to a coarse internal node, the leaves in the corresponding subtree are excluded from the standard loss and are neither explicitly encouraged nor discouraged individually. Instead, the $\operatorname{logsumexp}$ of their logits is computed, producing a single scalar value that smoothly approximates the maximum descendant leaf logit. This aggregated value is then encouraged to approach $1$. All leaves outside the subtree are still pushed toward $0$ as in the leaf-labeled case. A key structural consequence of this design is that, because each leaf prediction is produced via an independent sigmoid rather than a shared softmax, the leaf scores are never jointly normalized and therefore do not sum to one. Consequently, the score of a parent node has no principled relationship to the sum of the scores of its children, and the ratio of two sibling node scores carries no probabilistic meaning. The $\operatorname{logsumexp}$ aggregation used for coarse labels thus serves only as a heuristic scalar summary rather than representing a true probability of the internal node under a consistent probabilistic model.

In contrast, our approach imposes a single softmax over leaves, so that $\sum_{i \in \mathcal{L}} s_i = 1$ by construction, and defines each internal node's probability as the marginal probability mass over its descendant leaves, $p(v) = \sum_{l \in \text{leaves}(v)} s_l$. Since all node probabilities derive from the same normalized leaf distribution, they are directly comparable in magnitude and interpretable as marginal probabilities. We note, however, that in a DAG-structured ontology, where a leaf may have multiple parents, sibling internal nodes can share descendant leaves, meaning their marginal probabilities can sum to more than the parent's probability and the ratio $p(c) / p(v)$ can exceed $1$. Top-down decoding based on $p(c \mid v) = p(c) / p(v)$ is therefore a heuristic in this setting, as it is for any formulation applied to a DAG ontology. Nevertheless, our formulation retains a meaningful advantage: because all node probabilities derive from the same normalized leaf distribution via summation, they lie on a common probability scale and can be meaningfully compared across nodes at any level of the ontology, unlike the independent sigmoid scores of \textsc{scPRINT} which carry no such global meaning.

At inference, the ontology can be queried at any level of granularity. We support two decoding strategies.
\emph{Global leaf-argmax decoding} selects the single most probable leaf, $\hat{\ell} = \arg\max_{i \in \mathcal{L}} s_i$, and reports all of its ancestors as the predicted lineage.
\emph{Greedy hierarchical decoding} traverses the ontology from the root downward, selecting the child with the highest probability at each internal node:
\[
v_{t+1} = \arg\max_{u \in \mathrm{children}(v_t)} p(u),
\]
continuing until a leaf is reached.
The two strategies can disagree when probability mass is concentrated on a subtree whose individual leaves are each less probable than the global argmax leaf from a different subtree; greedy decoding favors the subtree with the largest aggregate mass, while leaf-argmax favors the single sharpest prediction.
In the results reported here, we use leaf-argmax decoding unless otherwise noted.

\subsection{Model Architecture}
The model architecture employs a bidi\-rectional, multi‑task adaptation of the LLaMA framework~\cite{Touvron2023LLaMA}, commonly used across diverse domains such as molecular conformation modeling~\cite{gurev2025standard} and time‑series analysis~\cite{rasul2023lag}. In our case, it is specifically configured to process continuous biological inputs. The primary architectural deviation from standard Large Language Models (LLMs) is the deliberate removal of the pre‑attention root mean square normali\-zation (RMSNorm)~\cite{Zhang2019RMSNorm} in the initial transformer block. Inspired by the AlphaFold 3 architecture~\cite{abramson2024accurate}, where coordinates are injected post‑normalization, this modification ensures that the absolute magnitudes of the floating‑point input representations are preserved rather than attenuated before their first attention operation.
Within the transformer layers, the network employs Swish‑Gated Linear Unit (SwiGLU) feed‑forward networks~\cite{Shazeer2020GLU} and omits internal dropout entirely to maintain representation stability, relying instead on weight decay for regularization. Dropout is applied exclusively in the terminal classification layers, which use Gaussian Error Linear Unit (GELU) activations~\cite{Hendrycks2016GELU}.

Models are implemented using Hugging Face Transformers with custom multifield tokenization for gene identity, expression values, and metadata. Training uses PyTorch Lightning for distributed execution, Hydra for configuration management, and TorchMetrics for evaluation. Loss functions include cross-entropy (classification), MSE and zero-inflated MSE (expression), and gradient reversal layers (batch correction). For interpretability, we employ integrated gradients~\cite{sundararajan_integrated_gradients_2017} via Captum~\cite{kokhlikyan2020captum} to compute gene-level attributions. Implementation details in Supplementary Methods Section 2.4.

\subsection{Pretraining Experiments}\label{checkpoints}

To illustrate the combined effects of WCED and hierarchical cell-type pretraining, we trained two multitask checkpoints:
WCED MULTITASK and MLM MULTITASK. Both checkpoints were trained on a 10\% downsampled version of CellXGene (2025-01-30 release) comprising ~6.2M cells, 19,362 protein coding genes only, based on HGNC. 
As discussed in Sec. \ref{sec:1pct-multitask}, the improvement when  moving from 1\% to 10\% of the data was marginal and we therefore did not train a checkpoint on the entire corpus.
Similar results have been recently reported by DenAdel et al.~\cite{denadel_evaluating_2024} and Kalfon et al.\cite{kalfon_scprint2_2025} indicating that training samples in scRNA can be downsampled significantly without loss of performance.
All of the data was log1p-normalized (count 10,000) and the task was expression prediction as a binary task (was expressed) and a quantification task (mean-squared error of log1p normalized level for gene). Training was conducted on 4x80GB A100 GPUs.
Based on the consistently inferior representations from MLM-only training (Section \ref{ref:mlmvswced}), we focused multitask pretraining resources on WCED and MLM multitask.

\subsubsection{WCED Multitask}
We found that warming up multitask training with a period of reconstruction only training improves performance, and so we began the multitask training after several epochs of WCED, and then continued pretraining until 8 epochs with cell type annotation using the hierarchical cross entropy loss, with a train val split by dataset (~5M samples in 394 datasets in train, ~1.2M samples in  84 in validation, selected so that all cell types are present during training and as many as possible in validation also, with any remaining datasets used for training).

\subsubsection{MLM Multitask}
Trained for 8 epochs with cell type annotation using the hierarchical cross entropy loss, with the same dataset split as the WCED Multitask checkpoint.

\subsection{Drug Mechanism of Action Prediction Task}

To evaluate the performance of different foundation model zero-shot representation for predicting drug mechanism of action (MoA), we utilized the Sci-Plex 3 A549 dataset containing single-cell transcriptomic profiles of A549 lung cancer cells treated with various compounds. To focus on cells exhibiting strong drug responses, we filtered to retain only cells with viability below 0.85 at 10,000 nM dose, as low viability indicates substantial cellular perturbation and more pronounced transcriptomic changes that better reflect the drug's mechanism of action. Vehicle controls and MoA categories with low numbers were excluded from analysis. From BMFM we used both both the MLM embeddings and WCED cell embedding with WCED using the reconstruction cell embeddings and the first CLS token representation. Additionally, we also used a concatenation of both these two embeddings. We compared embeddings from three state-of-the-art single-cell foundation models: Geneformer ~\cite{theodoris_transfer_2023}, scGPT ~\cite{cui_scgpt_2024} and Transcriptformer ~\cite{pearce_cross-species_2025}. We also evaluated a concatenated representation combining WCED with the first CLS token (CLS 0) from the transformer output. We employed stratified 2-fold cross-validation approach that was repeated 5 times with the following drug holdout strategy: each fold was constructed to ensure that at least one complete drug from each MoA class was held out in the test set. This approach prevents data leakage by ensuring the model never sees any cells from test drugs during training, thereby evaluating each model's ability to generalize to MoAs rather than memorizing drug-specific patterns. This approach and filtering left us with 9 MoA and 39 unique drugs. Within each fold we employed a probing classifier strategy where a logistic regression classifier was trained on each embeddings to predict MoA labels. The performance of the model was evaluated using AUC, Recall and F1-score. To evaluate how each embeddings preserved biological structure for the MoA task, we used the biological conservation metrics from the scib package. These included normalized mutual information (NMI), adjusted rand index (ARI) and average silhouette width (ASW), which were averaged to produce the avgBio score.

\begin{table*}[ht]
\centering
\caption{Detailed hyperparameters and training configuration for MLM Multitask and WCED Multitask models.}
\label{tab:hyperparams_detailed}
\resizebox{\textwidth}{!}{%
\begin{tabular}{@{}lcc@{}}
\toprule
\textbf{Category / Parameter} & \textbf{MLM Multitask} & \textbf{WCED Multitask} \\
\midrule
\multicolumn{3}{@{}l}{\textit{Architecture Specifications}} \\
\midrule
Base Model & LLaMA & LLaMA \\
Estimated Parameters & $\sim$32M & $\sim$47M \\
Hidden Size & 384 & 384 \\
Attention Heads & 12 & 12 \\
Hidden Layers & 12 & 12 \\
Intermediate Size & 1536  & 1536 \\
Activation Function & GELU & GELU \\
Dropout Probability & N/A & N/A \\
\midrule
\multicolumn{3}{@{}l}{\textit{Optimization \& Hardware Setup}} \\
\midrule
Hardware & \multicolumn{2}{c}{4 $\times$ NVIDIA A100 (80GB)} \\
Precision & \multicolumn{2}{c}{16-mixed (tf32\_mode: medium)} \\
Max Epochs & 10 & 10 \\
Learning Rate & $1 \times 10^{-4}$ & $1 \times 10^{-4}$ \\
Weight Decay & 0.10 & 0.10 \\
Batch Size (per GPU) & 8 & 8 \\
Gradient Accumulation Steps & 2 & 2 \\
\midrule
\multicolumn{3}{@{}l}{\textit{Data \& Vocabulary Dimensions}} \\
\midrule
Dataset & \multicolumn{2}{c}{CellXGene Nexus Index (10\% Random Split $\approx$ 5M Cells)} \\
Max Sequence Context & 8192 & 8192 \\
Zero-Expression Strategy & Interleave Ratio (0.9) & Interleave Ratio (0.9) \\
Modeling Strategy & Multitask & Multitask \\
\midrule
\multicolumn{3}{@{}l}{\textit{Masking \& Representation Strategy}} \\
\midrule
Masking Objective & MLM & WCED \\
Mask / Dropout Ratio & 95\% Masking & 20\% Sequence Dropout \\
Expression Encoder & Scale Adapt & Scale Adapt \\
\midrule
\multicolumn{3}{@{}l}{\textit{Loss Objectives \& Weightings}} \\
\midrule
Expression Loss Formulation & MSE \& Is-Zero BCE & MSE \& Is-Zero BCE \\
Cell Type Loss & HCE (wt: 1.0) & HCE (wt: 1.0) \\
Tissue Loss & Focal, $\gamma=2$ (wt: 1.0) & Focal, $\gamma=2$ (wt: 1.0) \\
Tissue General Loss & Focal, $\gamma=2$ (wt: 1.0) & Focal, $\gamma=2$ (wt: 1.0) \\
Donor ID Loss & Focal, $\gamma=2$ (wt: 0.05) & Focal, $\gamma=2$ (wt: 1.0) \\
\bottomrule
\end{tabular}%
}
\end{table*}

\section{Data Availability}
We pretrained on \cellxgene{}~\cite{program_cz_2025} accessed via TileDB-SOMA Census.
Data selection and preprocessing details in Supplementary Section~\ref{sec:si-software}.
We evaluated models using \scEval{} benchmark datasets~\cite{liu_evaluating_2024}, CZ Benchmarks datasets~\cite{cz-benchmarks}, and additional datasets used by scGPT~\cite{cui_scgpt_2024}, which provide standardized protocols for cell type clustering, batch integration, cell-type annotation, and other single-cell analysis tasks. Complete dataset descriptions in Supplementary Table~\ref{tab:datasets}.

\section{Code Availability}
All code for pretraining, fine-tuning and evaluation is available at \nopagebreak{\url{https://github.com/BiomedSciAI/biomed-multi-omic}}.

\section{Acknowledgments}
We would like to acknowledge Feixiong Cheng for numerous helpful discussions.

\section{Supplementary Material}

\input{supplement}

\end{document}

%% file: supplement.tex
\section{Gene Family Classification}
\label{sec:gene-family-classification}

To systematically analyze model performance across distinct biological
functions and expression programs, we partitioned the set of 19,027 genes
evaluated in our study along two orthogonal axes: a functional family
(what the gene does) and an expression program (how the gene is regulated).
Both axes produce mutually exclusive assignments; no model predictions or
training data were used to define either partition.

\subsection{Axis 1: Functional Family}
\label{sec:axis1-family}

\subsubsection{Classification strategy}
Because many genes possess pleiotropic functions and are annotated with
multiple Gene Ontology (GO) terms, a flat classification produces highly
overlapping groups.
To resolve this, we implemented a strict, priority-based hierarchical
partitioning scheme.
Genes were assigned to the first functional category they matched based
on a predefined biological hierarchy: hardware and identity (mitochondrial
genome, ribosomes, MHC), followed by transcriptional regulation,
structural components, cellular signaling, connectivity and adhesion,
receptors and transporters, cell-cycle and DNA-repair machinery,
homeostasis, and finally broad metabolic enzyme classes.
Genes that matched no category were labeled Unclassified.

\subsubsection{Annotation sources}
Gene annotations were derived from two primary sources:
\begin{enumerate}
    \item \textbf{The Gene Ontology (GO).}
    We utilized the human GO Annotation File (GAF) and the GO Directed
    Acyclic Graph (DAG; release 2025-10-10).
    For each category, genes were recursively matched not only to the
    root GO term but to all descendant terms within the DAG, ensuring
    comprehensive coverage.

    \item \textbf{TFClass Ontology.}
    The generic GO term for transcription factor activity
    (\texttt{GO:0003700}) encompasses over 1,400 genes.
    To achieve structural granularity, we incorporated the TFClass
    ontology and used RDF graph traversal to map each transcription factor
    to its top-level structural superclass (e.g., Zinc-coordinating,
    Helix-Turn-Helix, Basic domains).
    Historical synonyms and ortholog labels present in TFClass were
    programmatically normalized to match canonical HGNC symbols, yielding
    1,297 mapped genes across 10 superclasses.
\end{enumerate}

\subsubsection{Priority breakdown and yield}
Specific categories were refined to address edge cases.
The MHC/Antigen Presentation group was subdivided into Classical Class~I
(HLA-A, HLA-B, HLA-C, B2M), Class~II (HLA-D loci), Non-Classical
Class~I (HLA-E, HLA-F, HLA-G), and a residual antigen-presentation
category, enabling separate analysis of the distinct immunological roles
of each subclass.
The Cytoskeleton category was expanded to include Actin Binding
(\texttt{GO:0003779}) to capture structural genes such as \textit{TAGLN},
while Synapse and Cell Adhesion were split to distinguish neural
connectivity from general tissue integrity.
Cell Cycle (\texttt{GO:0007049}) and DNA Repair (\texttt{GO:0006281})
were assigned before the broad homeostasis categories (Ubiquitin Ligase,
Peptidase, RNA Binding) to prevent these biologically coherent programs
from being absorbed into larger enzyme or binding-protein groups.
The full priority order, defining criteria, and gene counts are detailed
in Supplementary Table~\ref{tab:gene_families}.

\vspace{1em}

\begin{longtable}{@{}r p{4.2cm} p{6.2cm} r@{}}
\caption{Gene Family Partitioning Hierarchy and Distribution}
\label{tab:gene_families} \\
\toprule
\textbf{Priority} & \textbf{Gene Family} & \textbf{Defining Criteria (GO Term / Rule)} & \textbf{Count} \\
\midrule
\endfirsthead

\multicolumn{4}{c}%
{{\bfseries \tablename\ \thetable{} -- continued from previous page}} \\
\toprule
\textbf{Priority} & \textbf{Gene Family} & \textbf{Defining Criteria (GO Term / Rule)} & \textbf{Count} \\
\midrule
\endhead

\midrule
\multicolumn{4}{r}{{Continued on next page}} \\
\endfoot

\bottomrule
\endlastfoot

\multicolumn{4}{l}{\textit{Hardware \& Identity}} \\
1 & Mitochondrial Genome & HGNC prefix: \texttt{MT-} & 13 \\
2 & Ribosome (Cytosolic) & \texttt{GO:0005840} $\cap$ prefixes \texttt{RPL, RPS, FAU, UBA52} & 91 \\
3 & Ribosome (Mitochondrial) & Prefixes: \texttt{MRPL, MRPS} & 77 \\
4a & MHC Class I (Classical) & \texttt{GO:0042611} $\cup$ prefix \texttt{HLA-}; HLA-A, -B, -C, B2M & 4 \\
4b & MHC Class II & Prefix: \texttt{HLA-D} & 13 \\
4c & MHC Class I (Non-Classical) & HLA-E, HLA-F, HLA-G & 3 \\
4d & MHC / Antigen Pres.\ (Other) & Remaining \texttt{GO:0042611} members & 3 \\
\addlinespace

\multicolumn{4}{l}{\textit{Transcriptional Regulation (via TFClass \& GO)}} \\
5 & TF: Zinc-Coordinating & TFClass superclass: Zinc-coordinating DNA-binding domains & 673 \\
6 & TF: Helix-Turn-Helix & TFClass superclass: Helix-turn-helix domains (homeodomains) & 362 \\
7 & TF: Basic Domains & TFClass superclass: Basic domains (bZIP, bHLH) & 139 \\
8 & TF: Other (Structural) & Remaining TFClass superclasses (p53, STAT, NF-$\kappa$B, etc.) & 116 \\
9 & TF: Unclassified & \texttt{GO:0003700} fallback for TFs not in TFClass & 281 \\
\addlinespace

\multicolumn{4}{l}{\textit{Structure \& Motility}} \\
10 & Cilia \& Flagella & \texttt{GO:0005929} & 510 \\
11 & Cytoskeleton \& Motor & \texttt{GO:0005856} $\cup$ \texttt{GO:0003774} $\cup$ \texttt{GO:0003779} & 1,082 \\
12 & Histone & Prefixes: \texttt{HIST, H2A, H2B, H3C, H4C} & 75 \\
\addlinespace

\multicolumn{4}{l}{\textit{Signaling \& Connectivity}} \\
13 & GTPase / Trafficking & \texttt{GO:0003924} & 271 \\
14 & Kinase & \texttt{GO:0016301} & 632 \\
15 & Phosphatase & \texttt{GO:0016791} & 229 \\
16 & Synapse & \texttt{GO:0045202} & 868 \\
17 & Cell Adhesion & \texttt{GO:0007155} & 539 \\
\addlinespace

\multicolumn{4}{l}{\textit{Receptors, Transporters \& Environment}} \\
18 & Cytokine / Growth Factor & \texttt{GO:0005125} $\cup$ \texttt{GO:0008083} & 284 \\
19 & Olfactory Receptor & Prefix: \texttt{OR} + digit & 403 \\
20 & GPCR (Non-Olfactory) & \texttt{GO:0004930} & 300 \\
21 & Ion Channel & \texttt{GO:0005216} & 287 \\
22 & Transporter (SLC/ABC) & \texttt{GO:0005215} $\cup$ prefix \texttt{SLC} & 714 \\
23 & Surface Marker (CD) & Prefix: \texttt{CD} + digit & 58 \\
\addlinespace

\multicolumn{4}{l}{\textit{Cell-Cycle \& Genome Maintenance}} \\
24 & Cell Cycle & \texttt{GO:0007049} & 167 \\
25 & DNA Repair & \texttt{GO:0006281} & 395 \\
\addlinespace

\multicolumn{4}{l}{\textit{Homeostasis \& Protein Processing}} \\
26 & Ubiquitin Ligase & \texttt{GO:0004842} & 350 \\
27 & Peptidase & \texttt{GO:0008233} & 488 \\
28 & Extracellular Matrix & \texttt{GO:0031012} & 203 \\
29 & RNA Binding & \texttt{GO:0003723} & 1,103 \\
\addlinespace

\multicolumn{4}{l}{\textit{Metabolism (Sub-classified Core Enzymes)}} \\
30 & Enzyme: Hydrolase & \texttt{GO:0016787} & 560 \\
31 & Enzyme: Transferase & \texttt{GO:0016740} & 756 \\
32 & Enzyme: Oxidoreductase & \texttt{GO:0016491} & 498 \\
33 & Enzyme: Other & \texttt{GO:0016829} $\cup$ \texttt{GO:0016874} $\cup$ \texttt{GO:0016853} & 226 \\
\addlinespace

\multicolumn{4}{l}{\textit{Remaining}} \\
34 & Unclassified & Genes not matching any of the above criteria & 6,254 \\
\midrule
\multicolumn{3}{r}{\textbf{Total Genes Classified}} & \textbf{19,027} \\

\end{longtable}

\subsection{Axis 2: Expression Regime}
\label{sec:axis2-regime}

The functional family axis groups genes by molecular role but does not
address whether a gene's expression is stable across cell states or
responsive to transient perturbations.
To capture this distinction, we introduced a second, orthogonal axis
assigning each gene to one of five expression programs derived entirely
from external databases and prior biological knowledge.

\subsubsection{External data sources}
Three public resources were combined:
\begin{enumerate}
    \item \textbf{Human Protein Atlas (HPA), v25}~\cite{uhlen2015tissue}.
    The ``RNA single cell type specificity'' annotation classifies genes
    by how narrowly their expression is distributed across cell types
    (20,151 genes with specificity data in the version used).
    Genes annotated as ``Cell type enriched'' or ``Group enriched'' were
    designated \emph{identity-associated}: their expression defines or
    distinguishes specific cell types.

    \item \textbf{Eisenberg \& Levanon housekeeping genes}~\cite{eisenberg2013human}.
    A curated set of 3,804 constitutively expressed genes identified
    across multiple human tissues via uniform expression criteria.

    \item \textbf{Dissociation-response genes.}
    Three sources were combined to capture genes whose expression changes
    as an artefact of tissue dissociation or acute cellular stress:
    (i)~the 15 core immediate early genes (IEGs) identified by van den
    Brink et~al.~\cite{vandenbrink2017single} as upregulated during
    enzymatic tissue dissociation (including \textit{FOS}, \textit{JUN},
    \textit{EGR1}, \textit{ATF3}, and \textit{HSPA1A/B});
    (ii)~genes annotated under ``response to heat''
    (\texttt{GO:0009408}; 89 genes); and
    (iii)~genes annotated under ``response to unfolded protein''
    (\texttt{GO:0006986}; 89 genes).
    The heat-response and UPR branches capture the broader stress
    transcriptome that co-activates during dissociation protocols.
\end{enumerate}

\subsubsection{Decision tree}
Each gene was assigned to exactly one expression program via a strict
priority rule applied in the following order:

\begin{enumerate}
    \item \textbf{\texttt{state\_dissociation}} (154 genes).
    Membership in the dissociation set (van den Brink core IEGs $\cup$
    GO heat response $\cup$ GO unfolded protein response), \emph{excluding}
    any gene that the HPA classifies as identity-associated.
    This exclusion prevents cell-type-specific stress genes from being
    misclassified as artefactual dissociation responders, following the
    logic of O'Flanagan et~al.~\cite{oflanagan2019dissociation}.

    \item \textbf{\texttt{state\_cell\_cycle}} (210 genes).
    Membership in the GO Cell Cycle branch (\texttt{GO:0007049}), again
    excluding HPA identity-associated genes.
    This captures proliferation-linked transcriptional programs that vary
    with mitotic phase rather than lineage.

    \item \textbf{\texttt{constitutive}} (3,338 genes; 17.5\%).
    Membership in the Eisenberg housekeeping set.
    These genes are expected to be expressed at relatively stable levels
    across cell types and conditions.

    \item \textbf{\texttt{identity\_associated}} (3,373 genes; 17.7\%).
    HPA ``Cell type enriched'' or ``Group enriched'' annotation, for
    genes not already assigned to a state category.

    \item \textbf{\texttt{unassigned}} (11,952 genes; 62.8\%).
    All remaining genes.
    This large residual reflects the limited coverage of existing
    annotation databases rather than a biological claim; many of these
    genes likely have context-dependent expression patterns that current
    resources do not capture.
\end{enumerate}

The priority order ensures that genes with documented state-responsive
behavior (dissociation artefacts, cell cycle) are not overridden by
identity or housekeeping labels.
Identity-associated genes are excluded from the two state categories
because a gene that is both cell-type-enriched and present in a stress
GO term (e.g., a lineage-specific heat shock protein) is more
informatively classified by its identity association for the purpose of
interpreting model learnability.

\subsubsection{Cross-tabulation with functional families}
The two axes are not independent.
RNA Binding genes are disproportionately constitutive (519 of 1,103;
47\%), consistent with their roles in ribosome biogenesis, mRNA
processing, and translation.
The Unclassified family contains the largest absolute number of
identity-associated genes (1,242 of 6,254; 20\%), reflecting the fact
that many tissue-specific genes lack well-curated GO annotations.
By contrast, Olfactory Receptor genes are almost entirely unassigned
(339 of 403; 84\%), with no housekeeping members and only 64
identity-associated genes, consistent with their highly restricted
expression in olfactory epithelium falling outside the HPA tissue panel.

\subsubsection{Rationale}
The expression program axis provides a principled, a~priori basis for
testing the hypothesis that model learnability reflects the biological
character of a gene's expression program.
If transcriptomic foundation models primarily learn stable cell-type
identity programs, identity-associated and constitutive genes should be
more learnable than state-responsive genes, independent of their
functional family.
The regime labels were fixed before any model output was examined,
preventing post hoc rationalization of performance differences.
\section{Analysis of winner and loser genes across model architectures}
\label{sec:supp-winners}

\paragraph{Outlier detection.}
We find that in general a gene's prediction error (rescaled MAE) is largely explained by the gene's sparsity.
The comparatively few non-sparse genes have significantly lower error and the error increases as the zero fraction increases.
This naturally raises the question as to whether or not there are genes that are significantly more learnable than others, 
for a given sparsity.
To assess this, we fit an isotonic regression trend to log-rescaled MAE as a function of zero fraction across all evaluated genes, enforcing a monotone non-decreasing relationship (sparser genes are harder to predict).
A local standard deviation was estimated within sliding windows of zero fraction to capture the heteroscedastic structure of the residuals (the spread of errors is itself larger for sparser genes).
Each gene's Z-score is its residual divided by this local standard deviation.
A gene was classified as a \emph{winner} (\emph{loser}) if $Z < -2$ ($Z >2$) and its rescaled MAE was more than 10\% below (above) the trend in absolute terms.
The practical significance filter prevents genes near the dense end of the distribution, where residuals are small in absolute terms, from qualifying on Z-score alone.
Classification was performed across the full sparsity range for both WCED (16,288 reliable genes evaluated) and MLM (7,327 genes evaluated); biological interpretation of winner and loser status focuses on the sparse regime (zero fraction $> 0.5$), as discussed below.
A gene was considered ``reliable'' if it appeared 100 times or more with a nonzero expression level.

\paragraph{Absolute and relative learnability.}
Two complementary measures of gene learnability emerge from this framework:
\emph{Absolute learnability} is the raw rescaled MAE or the fraction of genes beating the null baseline, which is the average nonzero value in the test set.
It is dominated by expression frequency: densely expressed genes (ribosomal proteins, mitochondrial-encoded genes) have low rescaled MAE because the model observes many non-zero training examples and learns a good estimate of the population mean.
\emph{Relative learnability} is measured by the sparsity-corrected Z-score, which asks whether a gene is predicted better or worse \emph{than other genes at the same expression frequency}.
Relative learnability is more surprising, indicating something about the gene beyond its frequency of appearance in the data that makes it easier to learn to predict the level for.

These two measures identify different gene populations as ``best'' in the quantification task.
Ribosomal proteins and mitochondrial-encoded genes achieve the lowest absolute MAE (0.10--0.15 rescaled) and the highest rates of beating the null baseline (80--85\%).
This is in line with expectations from biology where these genes are commonly expressed at high levels with low variability.
In contrast, the winner genes identified by the Z-score framework are predominantly sparse (median zero fraction 0.94 in WCED, 0.83 in MLM) and rarely beat the null baseline in absolute terms, yet their expression levels are significantly more predictable than other genes of comparable sparsity.
The biological insight lies almost entirely in this relative measure.

\paragraph{Z-score interpretability in the dense regime.}
The Z-score framework requires a sufficiently large and functionally
diverse local comparison group to produce interpretable outlier statistics.
Below approximately zero fraction 0.5, the gene population thins rapidly
and is dominated by a small number of constitutive families: cytosolic
ribosomal proteins (87 genes, median zero fraction 0.21), mitochondrial-encoded
genes (13 genes, median zero fraction 0.33), and scattered members of
RNA-binding and cytoskeletal families.
In this regime, the isotonic trend and local variance estimate are
determined by a narrow, functionally homogeneous population.
A ribosomal protein's Z-score reflects how it compares to other ribosomal
proteins and the handful of neighbouring constitutive genes, not to a
broad cross-section of the transcriptome.
Positive Z-scores for constitutive genes in this regime do not indicate
that these genes are ``hard to learn''; they indicate that the local
comparison group lacks the diversity needed for meaningful outlier
detection.
For this reason, although winner and loser classification is performed across the full sparsity range, we focus biological interpretation on the sparse regime (zero fraction $> 0.5$), where thousands of functionally diverse genes provide robust local statistics.

\paragraph{Winner and loser counts.}
Across the full sparsity range, WCED identified 909 winners and 6 losers; MLM identified 371 winners and 13 losers.
The vast majority of winners fall in the sparse regime (896 of 909 WCED winners and 361 of 371 MLM winners have zero fraction $> 0.5$), where biological interpretation is most reliable.
The large asymmetry between winners and losers stems from the fact that in the high zero fraction regime
most rescaled errors are high, approaching a ceiling at 1.
Rescaled MAE is effectively bounded at 1 in our analysis, and so variance is bounded in the direction of more error.
The small number of losers (6--13 genes) is not sufficient to support confident biological interpretation;
we list them for completeness in Table~\ref{tab:losers}.
All 15 unique loser genes have zero fraction below 0.65, placing them in the dense regime where Z-score
interpretability is limited (see above).

\paragraph{Cross-architecture agreement on winners.}
Of the 371 MLM winners, 283 (76.3\%) are also WCED winners.
Under the null hypothesis that winner status is independent across models,
the expected overlap is $\frac{909}{16{,}288} \times 371 \approx 21$ genes
(5.6\% of MLM winners); the observed 283 represents a 13.7-fold enrichment
($p < 10^{-30}$).
Among the 283 shared winners, the per-gene Z-scores are correlated across
models (Spearman $\rho = 0.66$, $p = 3.7 \times 10^{-37}$), confirming that
the degree of outperformance is consistent, not just the binary
winner/non-winner call.
Together these results establish that winner status reflects an intrinsic
biological property of each gene rather than a model-specific artefact.

The 626 WCED-exclusive winners are concentrated in synapse (94 genes),
cilia and flagella (53), and unclassified (183) families, biologically
indistinguishable from the shared winners.
Their WCED exclusivity reflects data volume rather than architecture:
cilia and flagella genes achieve an 11.4\% winner rate in WCED (58 of 510
annotated genes) but only 1.2\% in MLM (6 of 510), a 10-fold gap
consistent with ciliated epithelial cells being rare enough that
${\sim}$30-fold fewer training observations in MLM is insufficient to push
these genes past the significance threshold.
The 88 MLM-exclusive winners show no systematic family enrichment,
consistent with marginal cases tipping over threshold in one model but
not the other.

\paragraph{Family-level learnability.}
Gene families partition genes by biochemical function: what a protein does.
Within the sparse regime where relative learnability is interpretable,
winner rates reveal a clear hierarchy.

\textbf{High winner rate ($>$10\% of family genes):}
Surface Marker (CD), Synapse, Extracellular Matrix, Cilia \& Flagella,
Cell Adhesion, and Cytoskeleton \& Motor.
These families share a property: their member genes tend to be expressed
in narrow, stereotyped cell populations, where their expression levels
co-vary with broader cell-type programs that the model can read from
other genes in the profile.

\textbf{Moderate winner rate (3--8\%):}
Ion Channel, Transcription Factors (Basic Domains and Unclassified),
Phosphatase, Peptidase, Kinase, and Unclassified.

\textbf{Low winner rate ($<$3\%):}
GPCR, TF: Helix-Turn-Helix, Cytokine/Growth Factor, Ubiquitin Ligase,
RNA Binding, and TF: Zinc-Coordinating.
These families tend toward broad expression or encode programs modulated
by context rather than stable lineage.

\textbf{Dense regime (winner status not biologically interpretable):}
Ribosome (Cytosolic), Mitochondrial Genome, MHC/Antigen Presentation,
and Ribosome (Mitochondrial).
These families achieve the highest absolute performance (35--51\% median
MAE improvement, 80--85\% beating null), reflecting the straightforward
advantage of abundant training signal and low expression variance.
Although the outlier classification is applied to these genes (yielding the low constitutive winner rate reported below), their Z-scores reflect a narrow, functionally homogeneous comparison group rather than meaningful relative learnability (see above).

An instructive contrast is the CD surface marker family, which shows the
highest winner rate in both models (25.9\% WCED, 20.7\% MLM) yet 0\% of
CD genes beat the null model outright in WCED.
These results are not contradictory; they reflect the two different
measures.
CD genes are extremely sparse (median zero fraction 0.94), making their exact prediction difficult to learn.
But relative to other genes of the same sparsity, CD genes are systematically more predictable, because CD markers are among
the most reliable transcriptional correlates of immune cell identity.
This pattern, in which relative learnability dissociates from absolute performance, is the signature of genes whose expression levels are contextually predictable even when the gene itself is rarely observed.

\paragraph{Expression programs: testing contextual predictability.}
Gene families classify genes by biochemical function, but genes within the
same family can have very different expression patterns: a kinase can be
ubiquitously expressed or restricted to a single lineage.
To test what property of a gene's expression pattern drives relative
learnability, we classified genes into five expression programs based on
prior biological annotation, independently of model performance
(Supplementary Section~\ref{sec:axis2-regime}).

The expression program analysis directly tests whether contextual
predictability rather
than any particular biological function, determines relative learnability.
The prediction is that identity-associated genes, whose levels co-vary
with stable cell-type programs, should show excess learnability after
controlling for sparsity, while state programs whose levels are driven
by extra-transcriptomic factors (cell cycle phase, dissociation protocol)
should not.

The results support this prediction.
Identity-associated genes show significant excess learnability:
a 10.6\% winner rate in WCED (OR = 2.36 relative to unassigned,
$p = 7.1 \times 10^{-24}$) and 12.7\% in MLM (OR = 2.92,
$p = 5.3 \times 10^{-8}$), with Z-score medians shifted below the
unassigned baseline ($p = 2.0 \times 10^{-26}$ and
$p = 6.6 \times 10^{-3}$ respectively).
Constitutive genes are significantly \emph{depleted} among winners
(0.6\% winner rate, OR = 0.08, $p = 2.0 \times 10^{-65}$ in WCED),
consistent with the interpretive caveat above: their low sparsity places
them in a regime where Z-scores reflect the homogeneous comparison group,
not biological learnability.
Cell-cycle genes are also depleted (1.0\% winner rate, OR = 0.17,
$p = 1.6 \times 10^{-3}$ in WCED; 0.0\% in MLM, $p = 1.6 \times 10^{-2}$),
consistent with cell-cycle phase varying
\emph{within} cell types, so that cycle-gene levels are driven by mitotic
state rather than by the cell-type programs the model can read from context.

\paragraph{Dissociation-response genes.}
Dissociation-response genes present a conceptually important test case.
If the relevant axis were simply ``cell type vs.\ cell state,'' then
dissociation response, as a state program, should be unlearnable.
This prediction holds in the data: dissociation genes are not enriched
among winners (3.9\% winner rate, OR = 0.69, $p = 0.48$ in WCED;
3.7\%, OR = 0.73, $p = 0.66$ in MLM) and
their Z-score median is significantly higher (worse) than the unassigned
baseline ($p = 2.3 \times 10^{-3}$, WCED; $p = 3.1 \times 10^{-6}$, MLM).
The distinction between identity-associated and dissociation-response
genes is highly significant ($p = 1.4 \times 10^{-8}$, WCED;
$p = 1.5 \times 10^{-7}$, MLM).
Among the loser genes, HSP90AA1 and HSP90AB1 are both classified as
dissociation-response genes, consistent with the pattern, though the
small loser sample size precludes strong conclusions from this observation.

However, the simple ``type vs.\ state'' framing does not explain
\emph{why} dissociation genes' levels are not contextually predictable.
The key is that dissociation-gene expression levels are driven by factors
invisible to the model: the duration and severity of enzymatic digestion,
individual cell vulnerability, and thermal stress during
processing~\cite{vandenbrink2017single,oflanagan2019dissociation}.
These factors leave no systematic trace in the rest of the expression
profile, so the model has no basis for predicting the level from context.
The model can still detect \emph{whether} dissociation genes are expressed
(detection ROC-AUC is high for these genes) but cannot predict
\emph{how much} they are expressed, paralleling the broader
detection/quantification asymmetry we observe genome-wide.
We note this interpretation as a hypothesis consistent with the data,
not a conclusion.
What the data do establish is that dissociation-response genes are not
among the relatively learnable genes, despite being a well-characterized
transcriptional program with consistent co-activation structure.

The more precise characterization of relative learnability is therefore
not ``type genes are learnable and state genes are not,'' but rather:
\emph{a gene's expression level is contextually predictable when it co-varies with broader
transcriptional programs that the model can read from other genes in the
cell's profile.}
Cell-type identity programs are the principal source of such contextual
redundancy, which is why the simple type/state framing is a useful first
approximation.
This connects to recent findings that binary cell representations capture
much of the information in scRNA-seq data~\cite{bouland-nargb2021} and
that shuffling expression values among expressed genes preserves cell-type
embedding structure~\cite{moriel-neurips2025}.

\paragraph{Loser genes.}
A small number of genes were classified as losers
(6 WCED, 13 MLM; 4 shared: HSP90AA1, H3-3B, DDX5, RPS4Y1;
Table~\ref{tab:losers}).
These represent fewer than 0.1\% of evaluated genes, their
Z-scores are only modestly above the threshold, and all fall in the
dense regime (zero fraction $< 0.65$) where Z-score interpretability is
limited.
We list them for completeness but caution against over-interpreting
individual loser identities.

\begin{table}[p]
\centering
\caption{All loser genes ($Z > 2$, rescaled MAE $>10\%$ above trend)
identified by WCED and/or MLM.
Zero fraction (zf) is reported as a range where both models evaluated the
gene; single values indicate evaluation by one model only.
Dashes indicate the gene was not evaluated by that model.
All losers fall in the dense regime (zf $< 0.65$) where Z-score
interpretability is limited; individual identities should be interpreted
with caution.}
\label{tab:losers}
\footnotesize
\begin{tabular}{@{}lllcccc@{}}
\toprule
Gene & Family & Regime & zf &
\makebox[0.8cm]{WCED $Z$} & \makebox[0.8cm]{MLM $Z$} \\
\midrule
\multicolumn{6}{l}{\textit{Both models}} \\
HSP90AA1  & Cilia \& Flagella$^{*}$   & dissociation  & 0.17--0.22 & $+2.08$ & $+4.30$ \\
H3-3B     & Unclassified              & unassigned    & 0.14--0.19 & $+2.86$ & $+2.23$ \\
DDX5      & RNA Binding               & unassigned    & 0.19--0.24 & $+2.21$ & $+3.02$ \\
RPS4Y1    & Ribosome (Cytosolic)      & unassigned    & 0.58--0.64 & $+2.52$ & $+3.36$ \\
\addlinespace
\multicolumn{6}{l}{\textit{WCED only}} \\
HNRNPA2B1 & RNA Binding               & constitutive  & 0.27       & $+2.19$ & ---     \\
EIF1      & RNA Binding               & constitutive  & 0.16       & $+2.04$ & ---     \\
\addlinespace
\multicolumn{6}{l}{\textit{MLM only}} \\
CIRBP     & RNA Binding               & unassigned    & 0.28       & ---     & $+2.66$ \\
HMGB1     & Cytokine / Growth Factor  & constitutive  & 0.22       & ---     & $+2.51$ \\
PPIG      & RNA Binding               & constitutive  & 0.51       & ---     & $+2.33$ \\
CALM2     & Unclassified              & unassigned    & 0.26       & ---     & $+2.26$ \\
PRMT2     & Enzyme: Transferase       & unassigned    & 0.60       & ---     & $+2.22$ \\
ALDOA     & Cytoskeleton \& Motor     & unassigned    & 0.47       & ---     & $+2.11$ \\
MPHOSPH8  & Unclassified              & unassigned    & 0.56       & ---     & $+2.03$ \\
HSP90AB1  & RNA Binding               & dissociation  & 0.25       & ---     & $+2.02$ \\
HP1BP3    & Unclassified              & constitutive  & 0.52       & ---     & $+2.01$ \\
\bottomrule
\end{tabular}
\smallskip

\raggedright\footnotesize
$^{*}$HSP90AA1 is classified under Cilia \& Flagella in the v15 annotation
file; its biological function is as a stress-inducible chaperone
(HSP90$\alpha$).
\end{table}

\paragraph{Summary.}
By design, the winner classification controls for expression sparsity,
so enrichments among winners reflect properties of genes beyond how often
they are expressed.
The gene family hierarchy identifies \emph{which} families are enriched
among winners: those encoding cell-type-restricted structural and surface
proteins.
The expression program analysis identifies \emph{why}: genes annotated a
priori as identity-associated show significant excess learnability, while
genes annotated as constitutive, cell-cycle, or dissociation-response do
not.

The unifying principle is contextual predictability.
A gene's expression level is learnable beyond sparsity expectations when
it co-varies with broader transcriptional programs visible elsewhere in
the cell's profile, making it inferable from transcriptomic context.
Cell-type identity programs are the principal source of such contextual
redundancy.
This explains why cell-type annotation and batch integration
succeed as downstream tasks: they rely on the contextually predictable portion of the transcriptome.
However, gene-level expression prediction does not, because most genes' levels are driven by factors not recoverable from a single expression snapshot.

Absolute learnability, measured by raw MAE and null-beating rates, is
highest for densely expressed constitutive genes.
This reflects the advantage of abundant training signal and does
not require the model to learn cross-population structure.
Relative learnability, measured by sparsity-corrected Z-scores, identifies
the subset of the transcriptome where the model extracts biological signal
from co-expression patterns.
Both measures are valid; they answer different questions.
For understanding what TFMs learn about biology, relative learnability is
the more informative measure.

\begin{figure}[ht]
    \centering
    \includegraphics[width=\linewidth]{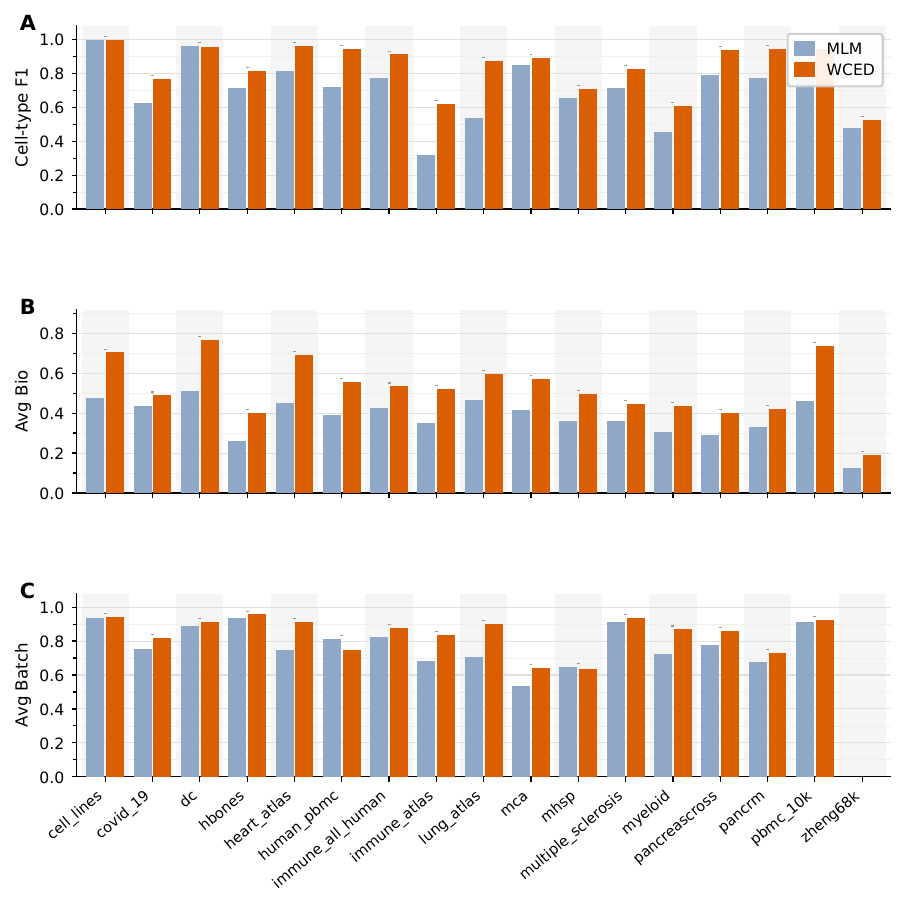}
    \caption{Zero-shot evaluation of \texttt{WCED} vs.\ \texttt{MLM} across 17 datasets (Table~\ref{tab:datasets}).
    Each panel shows paired bars (MLM in blue, WCED in orange) for
    (\textbf{A})~cell-type classification F1,
    (\textbf{B})~average biological conservation score, and
    (\textbf{C})~average batch integration score.
    WCED consistently outperforms MLM on biological conservation across all datasets,
    with substantial F1 gains on the majority of benchmarks.}
    \label{fig:wced_mlm_zero_shot}
\end{figure}

\begin{table}[ht]
\centering
\caption{Zero-shot evaluation: improvement of \texttt{WCED} over \texttt{MLM}
($\Delta = \texttt{WCED} - \texttt{MLM}$).
\textit{F1}: cell-type classification F1;
\textit{Avg Bio}: biological conservation;
\textit{Avg Batch}: batch integration.}
\label{tab:zero_shot_eval_full}
\small
\begin{tabular}{l r r r}
\toprule
& \multicolumn{3}{c}{$\Delta$ (\texttt{WCED} $-$ \texttt{MLM})} \\
\cmidrule(lr){2-4}
Dataset                      & {$\Delta$ F1} & {$\Delta$ Avg Bio} & {$\Delta$ Avg Batch} \\
\midrule
\texttt{cell\_lines}         & $-$0.001 & $+$0.227 & $+$0.006 \\
\texttt{covid\_19}           & $+$0.142 & $+$0.055 & $+$0.065 \\
\texttt{dc}                  & $-$0.003 & $+$0.256 & $+$0.023 \\
\texttt{hbones}              & $+$0.101 & $+$0.137 & $+$0.019 \\
\texttt{heart\_atlas}        & $+$0.150 & $+$0.242 & $+$0.163 \\
\texttt{human\_pbmc}         & $+$0.221 & $+$0.167 & $-$0.066 \\
\texttt{immune\_all\_human}  & $+$0.136 & $+$0.108 & $+$0.057 \\
\texttt{immune\_atlas}       & $+$0.299 & $+$0.171 & $+$0.156 \\
\texttt{lung\_atlas}         & $+$0.341 & $+$0.130 & $+$0.192 \\
\texttt{mca}                 & $+$0.042 & $+$0.154 & $+$0.104 \\
\texttt{mhsp}                & $+$0.053 & $+$0.135 & $-$0.015 \\
\texttt{multiple\_sclerosis} & $+$0.111 & $+$0.086 & $+$0.024 \\
\texttt{myeloid}             & $+$0.154 & $+$0.128 & $+$0.148 \\
\texttt{pancreascross}       & $+$0.148 & $+$0.113 & $+$0.082 \\
\texttt{pancrm}              & $+$0.171 & $+$0.088 & $+$0.055 \\
\texttt{pbmc\_10k}           & $+$0.067 & $+$0.276 & $+$0.016 \\
\texttt{zheng68k}            & $+$0.044 & $+$0.063 & ---      \\
\midrule
\textbf{Average}             & $\mathbf{+0.128}$ & $\mathbf{+0.149}$ & $\mathbf{+0.064}$ \\
\bottomrule
\end{tabular}
\end{table}



\begin{table}[htbp]
\centering
\caption{Weighted Combination of Average Bio (60\%) and Batch (40\%) for BMFM models and scGPT across scEval datasets}
\label{tab:Weighted_BIO_Batch}
\begin{tabular}{lp{2cm}p{2cm}p{2cm}p{2cm}}
\toprule
Method & BMFM \newline CONCAT & MLM \newline MULTITASK & WCED \newline MULTITASK & scGPT \\
Dataset &  &  &  &  \\
\midrule
cell\_lines & 0.801 & 0.793 & \textbf{0.819} & 0.813 \\
covid\_19 & 0.620 & 0.597 & 0.625 & \textbf{0.636} \\
dc & 0.847 & \textbf{0.849} & 0.844 & 0.820 \\
heart\_atlas & 0.792 & 0.795 & \textbf{0.802} & 0.801 \\
human\_pbmc & 0.701 & \textbf{0.718} & 0.686 & 0.672 \\
immune\_all\_human & 0.722 & \textbf{0.735} & 0.719 & 0.681 \\
immune\_atlas & 0.722 & \textbf{0.745} & 0.718 & 0.696 \\
lung\_atlas & 0.706 & 0.722 & 0.711 & \textbf{0.756} \\
pancrm & 0.744 & 0.721 & \textbf{0.752} & 0.564 \\
pbmc\_10k & 0.809 & 0.796 & 0.805 & \textbf{0.859} \\
\hline
\textbf{Average} & 0.746 & 0.747 & \textbf{0.748} & 0.730 \\
\bottomrule
\end{tabular}
\end{table}
\begin{table}[htbp]
\centering
\caption{Zeroshot average bio scores of BMFM models and scGPT across scEval datasets}
\label{tab:Average_Bio}
\begin{tabular}{lp{2cm}p{2cm}p{2cm}p{2cm}}
\toprule
Method & BMFM \newline CONCAT & MLM \newline MULTITASK & WCED \newline MULTITASK & scGPT \\
Dataset &  &  &  &  \\
\midrule
cell\_lines & 0.709 & 0.710 & \textbf{0.725} & 0.721 \\
covid\_19 & 0.469 & 0.455 & 0.470 & \textbf{0.486} \\
dc & \textbf{0.774} & \textbf{0.774} & 0.772 & 0.756 \\
heart\_atlas & 0.685 & 0.696 & 0.710 & \textbf{0.726} \\
human\_pbmc & 0.611 & \textbf{0.624} & 0.606 & 0.547 \\
immune\_all\_human & 0.588 & \textbf{0.616} & 0.589 & 0.533 \\
immune\_atlas & 0.616 & \textbf{0.659} & 0.616 & 0.596 \\
lung\_atlas & 0.565 & 0.597 & 0.563 & \textbf{0.661} \\
pancrm & 0.638 & 0.615 & \textbf{0.647} & 0.450 \\
pbmc\_10k & 0.698 & 0.696 & 0.694 & \textbf{0.800} \\
\bottomrule
\end{tabular}
\end{table}

\begin{table}[htbp]
\centering
\caption{Zeroshot average batch scores of BMFM models and scGPT across scEval datasets}
\label{tab:Average_Batch}
\begin{tabular}{lp{2cm}p{2cm}p{2cm}p{2cm}}
\toprule
Method & BMFM \newline CONCAT & MLM \newline MULTITASK & WCED \newline MULTITASK & scGPT \\
Dataset &  &  &  &  \\
\midrule
cell\_lines & 0.938 & 0.917 & \textbf{0.959} & 0.951 \\
covid\_19 & 0.847 & 0.811 & 0.857 & \textbf{0.861} \\
dc & 0.957 & \textbf{0.961} & 0.953 & 0.916 \\
heart\_atlas & \textbf{0.952} & 0.944 & 0.941 & 0.914 \\
human\_pbmc & 0.836 & \textbf{0.860} & 0.805 & \textbf{0.860} \\
immune\_all\_human & \textbf{0.924} & 0.914 & 0.915 & 0.903 \\
immune\_atlas & \textbf{0.882} & 0.874 & 0.872 & 0.847 \\
lung\_atlas & 0.918 & 0.910 & \textbf{0.933} & 0.898 \\
pancrm & 0.903 & 0.880 & \textbf{0.910} & 0.735 \\
pbmc\_10k & \textbf{0.975} & 0.947 & 0.972 & 0.948 \\
\bottomrule
\end{tabular}
\end{table}

\begin{table}[htbp]
\centering
\small
\label{tab:mouse_zeroshot_combined_metrics}
\caption{Zero-shot Performance for scEval mouse datasets: Average Bio, Average Batch, and Combined Score (0.6×Bio + 0.4×Batch)}
\begin{tabular}{p{1.5cm}p{0.95cm}@{\hspace{0.4cm}}p{0.95cm}@{\hspace{0.4cm}}p{0.95cm}@{\hspace{0.4cm}}p{0.95cm}@{\hspace{0.4cm}}p{0.95cm}@{\hspace{0.4cm}}p{0.95cm}}
\toprule
Metric & \multicolumn{2}{c}{Average Bio} & \multicolumn{2}{c}{Average Batch} & \multicolumn{2}{c}{Combined Score} \\
\cmidrule(lr){2-3} \cmidrule(lr){4-5} \cmidrule(lr){6-7}
Dataset & \small {MLM\newline MT} & \small {WCED\newline MT} & \small {MLM\newline MT} & \small {WCED\newline MT} & \small {MLM\newline MT} & \small {WCED\newline MT} \\
\midrule
mca & 0.583 & \textbf{0.628} & 0.867 & \textbf{0.894} & 0.697 & \textbf{0.734} \\
mhsp & \textbf{0.526} & 0.524 & 0.784 & \textbf{0.863} & 0.629 & \textbf{0.660} \\
\bottomrule
\end{tabular}
\end{table}
\begin{table*}[htbp]
\centering
\small
\caption{CZ Benchmarks Cell Type Classification Task - Macro F1 of models across datasets}

\label{tab:czi_f1}
\begin{tabular}{l@{\hspace{0.3cm}}c@{\hspace{0.3cm}}c@{\hspace{0.3cm}}c@{\hspace{0.3cm}}c@{\hspace{0.3cm}}c@{\hspace{0.3cm}}c}
\toprule
Model & MEAN & Blood & Bone & Lung & Mammary & Thymus \\
 & (all datasets) &  & marrow &  &  &  \\
\midrule
AIDO.Cell: 3M & 0.762 & 0.768 & 0.650 & 0.753 & 0.872 & 0.766 \\
BMFM-RNA-CONCATENATION & 0.825 & 0.812 & \textbf{0.730} & 0.813 & \textbf{0.900} & 0.871 \\
BMFM-RNA-MLM-MULTITASK & 0.806 & 0.797 & 0.708 & 0.794 & 0.887 & 0.844 \\
BMFM-RNA-WCED-MULTITASK & 0.816 & 0.807 & 0.715 & 0.803 & 0.888 & 0.867 \\
Baseline & 0.559 & 0.531 & 0.441 & 0.583 & 0.650 & 0.590 \\
Geneformer: 316M & 0.798 & 0.787 & 0.667 & 0.796 & 0.890 & 0.849 \\
TF-Exemplar & \textbf{0.828} & \textbf{0.813} & 0.707 & 0.829 & 0.895 & \textbf{0.895} \\
TF-Metazoa & 0.827 & 0.813 & 0.706 & \textbf{0.831} & 0.896 & 0.891 \\
TF-Sapiens & 0.825 & 0.811 & 0.703 & 0.825 & 0.894 & 0.895 \\
UCE: 33L & 0.808 & 0.776 & 0.684 & 0.812 & 0.886 & 0.885 \\
UCE: 4L & 0.806 & 0.780 & 0.680 & 0.819 & 0.880 & 0.869 \\
scGPT: whole-human & 0.810 & 0.794 & 0.704 & 0.819 & 0.880 & 0.851 \\
scVI: homo\_sapiens & 0.795 & 0.781 & 0.693 & 0.801 & 0.884 & 0.818 \\
\bottomrule
\end{tabular}
\end{table*}

\begin{table*}[htbp]
\centering
\small
\caption{CZ Benchmarks Cell Type Classification Task - Accuracy of models across datasets}
\label{tab:czi_accuracy}
\begin{tabular}{l@{\hspace{0.3cm}}c@{\hspace{0.3cm}}c@{\hspace{0.3cm}}c@{\hspace{0.3cm}}c@{\hspace{0.3cm}}c@{\hspace{0.3cm}}c}
\toprule
Model & MEAN & Blood & Bone & Lung & Mammary & Thymus \\
 & (all datasets) &  & marrow &  &  &  \\
\midrule
AIDO.Cell: 3M & 0.891 & 0.890 & 0.847 & 0.899 & 0.990 & 0.827 \\
BMFM-RNA-CONCATENATION & 0.920 & 0.906 & \textbf{0.872} & 0.937 & 0.994 & 0.891 \\
BMFM-RNA-MLM-MULTITASK & 0.909 & 0.897 & 0.860 & 0.928 & 0.993 & 0.869 \\
BMFM-RNA-WCED-MULTITASK & 0.917 & 0.905 & 0.866 & 0.932 & 0.994 & 0.888 \\
Baseline & 0.697 & 0.643 & 0.626 & 0.728 & 0.862 & 0.628 \\
Geneformer: 316M & 0.913 & 0.898 & 0.858 & 0.932 & 0.993 & 0.885 \\
TF-Exemplar & 0.925 & \textbf{0.908} & 0.869 & 0.943 & 0.994 & 0.913 \\
TF-Metazoa & 0.924 & 0.906 & 0.870 & 0.942 & \textbf{0.994} & 0.910 \\
TF-Sapiens & \textbf{0.926} & 0.907 & 0.870 & \textbf{0.943} & 0.994 & \textbf{0.914} \\
UCE: 33L & 0.912 & 0.886 & 0.852 & 0.930 & 0.994 & 0.898 \\
UCE: 4L & 0.908 & 0.883 & 0.850 & 0.930 & 0.993 & 0.886 \\
scGPT: whole-human & 0.912 & 0.892 & 0.866 & 0.933 & 0.993 & 0.877 \\
scVI: homo\_sapiens & 0.905 & 0.887 & 0.862 & 0.927 & 0.994 & 0.858 \\
\bottomrule
\end{tabular}
\end{table*}

\begin{table*}[htbp]
\centering
\small
\caption{CZ Cell Type Cell Clustering Task - ARI of models across datasets}
\label{tab:czi_cell_clustering_ari}
\begin{tabular}{l@{\hspace{0.3cm}}c@{\hspace{0.3cm}}c@{\hspace{0.3cm}}c@{\hspace{0.3cm}}c@{\hspace{0.3cm}}c}
\toprule
Model & Blood & Bone & Lung & Mammary & Thymus \\
 &  & marrow &  &  &  \\
\midrule
Baseline & 0.352 & 0.285 & 0.643 & 0.291 & 0.439 \\
AIDO.Cell: 3M & 0.229 & 0.313 & 0.418 & 0.166 & 0.245 \\
BMFM-RNA-CONCATENATION & 0.407 & 0.391 & 0.683 & \textbf{0.419} & 0.414 \\
BMFM-RNA-MLM-MULTITASK & 0.345 & 0.391 & \textbf{0.701} & 0.315 & 0.402 \\
BMFM-RNA-WCED-MULTITASK & 0.408 & \textbf{0.407} & 0.696 & 0.407 & 0.415 \\
Geneformer: 316M & 0.258 & 0.298 & 0.513 & 0.217 & 0.290 \\
TF-Exemplar & 0.274 & 0.347 & 0.657 & 0.244 & 0.421 \\
TF-Metazoa & 0.262 & 0.339 & 0.648 & 0.244 & \textbf{0.456} \\
TF-Sapiens & 0.292 & 0.344 & 0.665 & 0.245 & 0.451 \\
UCE: 33L & 0.298 & 0.371 & 0.536 & 0.293 & 0.428 \\
UCE: 4L & \textbf{0.416} & 0.394 & 0.648 & 0.341 & 0.432 \\
scGPT: whole-human & 0.294 & 0.367 & 0.592 & 0.269 & 0.316 \\
scVI: homo\_sapiens & 0.315 & 0.402 & 0.669 & 0.338 & 0.410 \\
\bottomrule
\end{tabular}
\end{table*}

\begin{table*}[htbp]
\centering
\small
\caption{CZ Cell Type Cell Clustering Task - NMI of models across datasets}
\label{tab:czi_cell_clustering_nmi}
\begin{tabular}{l@{\hspace{0.3cm}}c@{\hspace{0.3cm}}c@{\hspace{0.3cm}}c@{\hspace{0.3cm}}c@{\hspace{0.3cm}}c}
\toprule
Model & Blood & Bone & Lung & Mammary & Thymus \\
 &  & marrow &  &  &  \\
\midrule
Baseline & 0.661 & 0.653 & 0.830 & 0.703 & 0.719 \\
AIDO.Cell: 3M & 0.592 & 0.605 & 0.707 & 0.607 & 0.576 \\
BMFM-RNA-CONCATENATION & 0.701 & 0.694 & 0.837 & \textbf{0.781} & 0.702 \\
BMFM-RNA-MLM-MULTITASK & 0.676 & 0.687 & 0.836 & 0.725 & 0.690 \\
BMFM-RNA-WCED-MULTITASK & \textbf{0.703} & \textbf{0.698} & \textbf{0.842} & 0.774 & 0.704 \\
Geneformer: 316M & 0.612 & 0.606 & 0.770 & 0.648 & 0.610 \\
TF-Exemplar & 0.643 & 0.647 & 0.826 & 0.681 & 0.683 \\
TF-Metazoa & 0.639 & 0.651 & 0.824 & 0.679 & 0.689 \\
TF-Sapiens & 0.649 & 0.649 & 0.825 & 0.681 & 0.691 \\
UCE: 33L & 0.648 & 0.686 & 0.796 & 0.712 & \textbf{0.721} \\
UCE: 4L & 0.680 & 0.695 & 0.824 & 0.741 & 0.714 \\
scGPT: whole-human & 0.644 & 0.672 & 0.805 & 0.699 & 0.653 \\
scVI: homo\_sapiens & 0.652 & 0.679 & 0.823 & 0.737 & 0.665 \\
\bottomrule
\end{tabular}
\end{table*}

\begin{table*}[htbp]
\centering
\small
\caption{CZ Cell Type Cell Clustering Task - Silhouette Score of models across datasets}
\label{tab:czi_cell_clustering_ss}
\begin{tabular}{l@{\hspace{0.3cm}}c@{\hspace{0.3cm}}c@{\hspace{0.3cm}}c@{\hspace{0.3cm}}c@{\hspace{0.3cm}}c}
\toprule
Model & Blood & Bone & Lung & Mammary & Thymus \\
 &  & marrow &  &  &  \\
\midrule
AIDO.Cell: 3M & 0.635 & 0.457 & 0.490 & 0.502 & 0.484 \\
BMFM-RNA-CONCATENATION & 0.628 & 0.552 & 0.609 & 0.664 & 0.559 \\
BMFM-RNA-MLM-MULTITASK & 0.617 & 0.538 & \textbf{0.618} & \textbf{0.696} & 0.541 \\
BMFM-RNA-WCED-MULTITASK & 0.629 & \textbf{0.552} & 0.608 & 0.662 & 0.560 \\
Geneformer: 316M & 0.598 & 0.504 & 0.525 & 0.550 & 0.521 \\
TF-Exemplar & 0.614 & 0.526 & 0.589 & 0.589 & 0.552 \\
TF-Metazoa & 0.628 & 0.528 & 0.593 & 0.592 & 0.553 \\
TF-Sapiens & 0.608 & 0.522 & 0.585 & 0.581 & 0.548 \\
UCE: 33L & 0.651 & 0.544 & 0.601 & 0.628 & \textbf{0.571} \\
UCE: 4L & \textbf{0.670} & 0.543 & 0.615 & 0.637 & 0.564 \\
scGPT: whole-human & 0.647 & 0.538 & 0.568 & 0.595 & 0.539 \\
scVI: homo\_sapiens & 0.655 & 0.545 & 0.586 & 0.646 & 0.546 \\
\bottomrule
\end{tabular}
\end{table*}

\section{Identifying Disease-Associated Subtypes in Alzheimer's Disease}
\label{sec:alzheimers-daa}

We applied the BMFM-RNA WCED model to the Alzheimer's disease snRNA-seq dataset GSE138852 \cite{grubman_single-cell_2019} (Table~\ref{tab:datasets}; 13,214 nuclei from 6 AD and 6 control brains) to determine if the model could identify the same distinct set of disease-associated astrocytes (DAAs) as reported in previous work \cite{grubman_single-cell_2019, xu_graph_2023}.  In the resulting embedding, astrocytes clustered into multiple subpopulations, including a distinct cluster with the hallmark transcriptional signature of DAAs: elevated GFAP, CD44, HSPB1, TNC alongside reduced SLC1A2, SLC1A3, GLUL, NRXN1, CADM2, PTN, GPC5, which is consistent with previously validated DAA markers and the DAA cluster reported in the original analysis \cite{xu_graph_2023}.

\begin{figure}[htbp]
    \centering
    \begin{minipage}{0.49\linewidth}
        \centering
        \includegraphics[height=4cm]{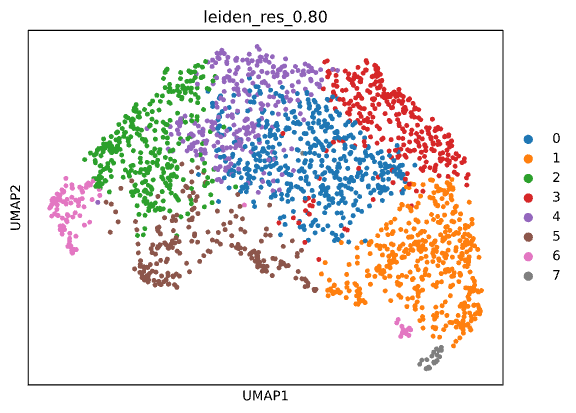}
    \end{minipage}
    \hfill
    \begin{minipage}{0.49\linewidth}
        \centering
        \includegraphics[height=5cm]{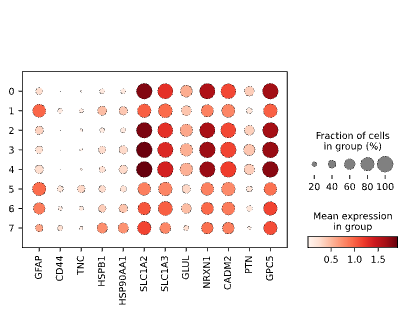}
    \end{minipage}
    \caption{(A) BMFM embeddings projected as a UMAP for GSE138852. (B) Dot plot of canonical DAA marker expression across astrocyte subclusters. Dot size indicates the fraction of cells expressing each gene and color intensity indicates scaled average expression.}
    \label{fig:astrocyte}
\end{figure}

\section{The \biomedmultiomic{} Software Architecture}
\label{sec:si-software}

Beyond the specific scientific evaluations presented in the main text, the \bmfmrna{} repository (\url{https://github.com/BiomedSciAI/biomed-multi-omic}) constitutes a comprehensive, modular software engine for building and evaluating multi-omic foundation models.
The framework unifies diverse data-streaming strategies, tokenization schemes, model architectures, pretraining objectives, and evaluation protocols within a single composable codebase.
Here we describe the software architecture in detail, complementing the compact overview in Methods.

\subsection{Framework Design Philosophy}

The software is organized around three engineering principles: modularity, scalability, and reproducibility.

\subsubsection{Modularity and composability}
Every component from data loading and tokenization to model architecture and training objective is independently configurable via YAML files using the Hydra configuration framework~\cite{yadan2019hydra}.
This allows researchers to systematically vary individual design choices without modifying the underlying training loops.
For example, transitioning a model from masked language modeling (MLM) to whole cell expression decoding (WCED), activating a multitask head to simultaneously predict hierarchical cell types, or replacing a standard transformer encoder with a linear-complexity alternative each requires only configuration overrides.

\subsubsection{Scalability}
The framework is designed to handle atlas-scale datasets natively, bypassing local memory constraints through several mechanisms:
(i)~TileDB-SOMA integration for cloud-native, out-of-core access to the CZ \cellxgene{} Census, eliminating the need for local copies of multi-billion-observation corpora;
(ii)~LitData streaming arrays for asynchronous, memory-efficient distributed data loading that decouples I/O from computation;
(iii)~PyTorch Lightning orchestration for automatic multi-node, multi-GPU distribution with minimal boilerplate; and
(iv)~mixed-precision training (FP16/BF16) coupled with gradient checkpointing to reduce memory footprint per device.
Together, these allow training runs on the full \cellxgene{} corpus (${\sim}62$M cells at the time of writing) to proceed on modest GPU allocations (4$\times$A100 80\,GB for the experiments reported here).

\subsubsection{Reproducibility}
All experimental pipelines enforce provenance tracking through:
(i)~versioned Hydra configurations automatically logged via ClearML, ensuring that every hyperparameter, data path, and architectural choice is recorded;
(ii)~deterministic data sampling pipelines with configurable, cross-node random seed management; and
(iii)~automated checkpoint management with distributed state recovery, enabling fault-tolerant training on preemptible infrastructure.
Metrics are computed via TorchMetrics in a distributed-safe manner, guaranteeing consistent evaluation across varying numbers of GPUs.

\begin{figure}[H]
    \centering
    \includegraphics[width=\linewidth]{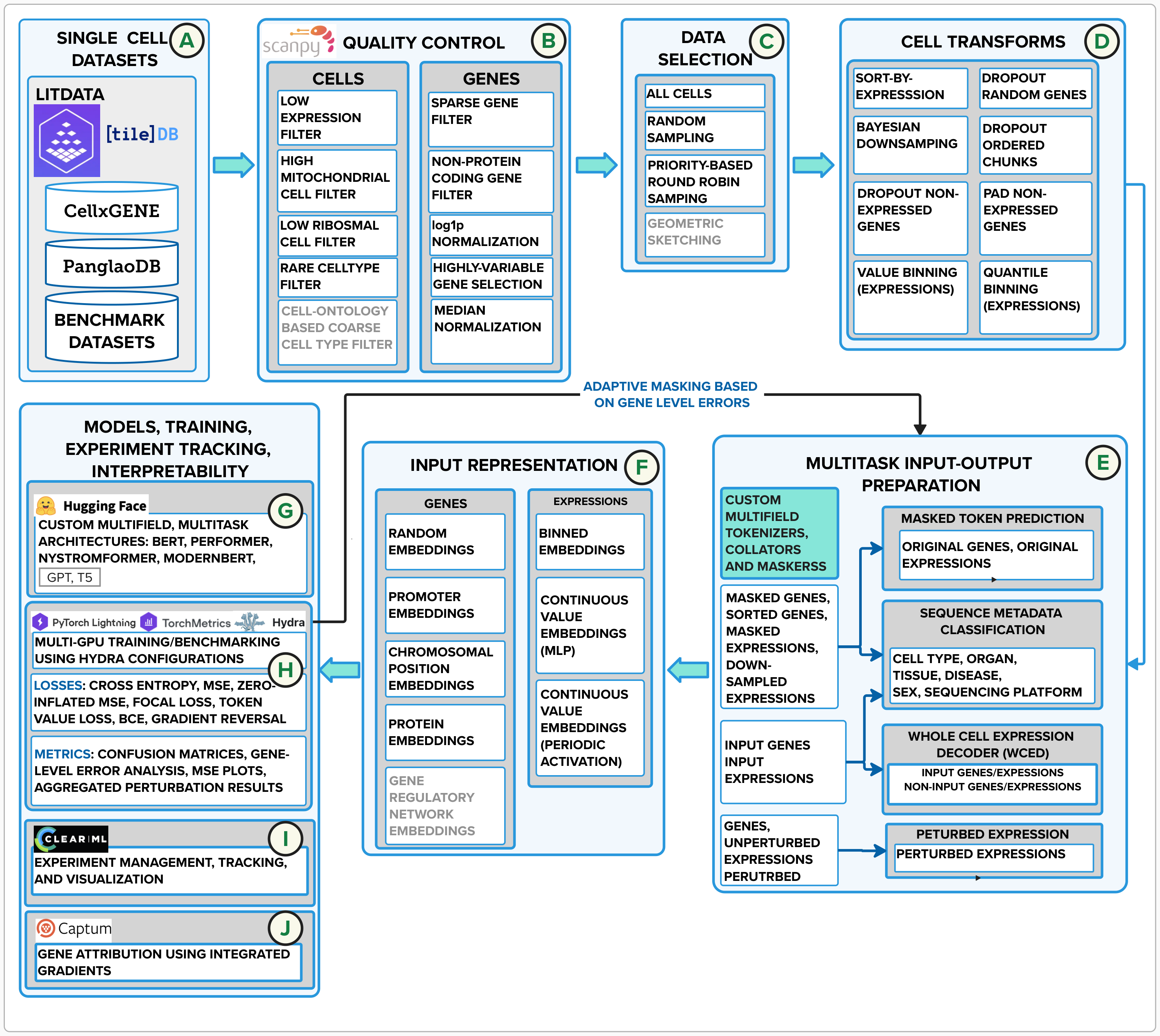}
    \caption{\bmfmrna{} framework diagram.
    \textbf{A.} scRNA data can be read from TileDB-SOMA (eg \cellxgene{}) or h5ad, and processed with LitData to support multi-gpu training.
    \textbf{B.} Quality control pipeline built on top of Scanpy.
    \textbf{C.} Data selection techniques to prioritize informative samples.
    \textbf{D.} Sample transformations/augmentations including read-depth aware down-sampling, non-expressed gene sampling, expression binning, gene ordering and dropout.
    \textbf{E.} Multitask input preparation supports masked language modeling on gene and/or expression inputs, sequence classification across multiple label types (e.g., cell type), joint training of both tasks, and sequence labeling for perturbed expression prediction.
    \textbf{F.} Model input representation supports integration of external gene-level knowledge, with expression values represented as either discrete tokens or continuous embeddings.
    \textbf{G.} Customized Transformer architectures based on HuggingFace including \BERT{}, \ModernBERT{}, \Performer{} and \Nystromformer{} to support multi-field inputs (genes, expressions, perturbation state) and multi-task learning.
    \textbf{H.} Training is implemented with PyTorch Lightning and fully configurable via Hydra, supporting flexible benchmarking. Multi-task training uses weighted losses built with TorchMetrics, including cross-entropy, focal loss, and zero-inflated MSE. Gradient reversal is incorporated to mitigate batch effects.
    \textbf{I.} ClearML is used for seamless experiment tracking and interactive visualization of results.
    \textbf{J.} Captum is used for model interpretability, allowing gene-level attributions to highlight important features driving predictions.
    \textbf{*} Orange indicates in-development.
    }
    \label{fig:si-package-overview}
\end{figure}

\paragraph{Scanpy and AnnData object integration}

The package provides a simple scanpy-style API for zero-shot inference on single-cell RNA-seq data. The \texttt{bmfm.inference()} function generates cell embeddings and predictions directly from pre-trained models held on HuggingFace. The results are then stored in the AnnData object allowing users to integrate with their existing workflows. The \texttt{pooling\_method} parameter controls how cell embeddings are extracted: \texttt{"first\_token"} uses the CLS token, \texttt{"mean\_pooling"} averages all tokens, \texttt{"pooling\_layer"} uses a trained pooler (if available) or a list concatenates multiple methods. Additional parameters include \texttt{layer} (AnnData layer to use), \texttt{batch\_size}, \texttt{device} and \texttt{checkpoint} (model path or HuggingFace identifier).

\begin{lstlisting}[language=Python, frame=single, numbers=left, basicstyle=\small\ttfamily, backgroundcolor=\color{lightgray!10}]
import bmfm_targets as bmfm
import scanpy as sc

# Load data and run inference
adata = sc.read_h5ad("data.h5ad")
adata = bmfm.inference(adata)

# Embeddings stored in adata.obsm['X_bmfm']
# Predictions stored in adata.obs['bmfm_pred_*']

# Use different pooling methods
adata = bmfm.inference(adata, pooling_method="mean_pooling")
adata = bmfm.inference(
    adata,
    pooling_method=["first_token", "mean_pooling"]
)
\end{lstlisting}

\subsection{Atlas-Scale Data Pipelines}

\subsubsection{Data access}
The framework provides unified data-loading interfaces for both local files (AnnData/h5ad) and cloud-hosted stores (TileDB-SOMA).
TileDB-SOMA access enables direct streaming from the CZ \cellxgene{} Census without requiring local downloads or format conversions.
For local workflows, standard AnnData objects are loaded via memory-mapped access where possible.
Both pathways feed into a common cell-level iterator, ensuring that downstream tokenization and batching logic is agnostic to the storage backend.

\subsubsection{Quality control}
Preprocessing follows standard single-cell quality control protocols~\cite{wolf-scanpy-2018}.
Cell-level filters remove observations with anomalously low unique molecular identifier (UMI) counts or high mitochondrial transcript fractions.
Gene-level filters retain only genes exceeding configurable thresholds for minimum expression frequency and dispersion.
All filtering thresholds are specified in the Hydra configuration and logged for reproducibility.
Expression values are then normalized per cell (total count normalization to a target sum of 10{,}000) and log-transformed via $\log(1+x)$.

\subsubsection{Sampling strategies}
Public single-cell atlases exhibit extreme class imbalance: a small number of tissues, donors, and cell types contribute the majority of observations, while rare populations may comprise fewer than 100 cells.
Na\"ive uniform sampling during pretraining would therefore overwhelm the model with redundant examples from dominant classes.
To address this, the framework implements configurable sampling strategies operating at multiple levels of the data hierarchy:
(i)~\emph{equal downsampling}, which caps the maximum number of cells drawn from any single dataset, tissue, or cell type;
(ii)~\emph{round-robin sampling}, which cycles through datasets or metadata categories in balanced fashion during each epoch; and
(iii)~\emph{label-aware stratification}, which ensures that user-specified metadata fields (e.g., cell type, disease status) are represented at controlled frequencies.
These strategies are composable and their parameters (cap sizes, field priorities, random seeds) are fully specified via configuration.

\subsection{Multifield Tokenization}

Standard NLP tokenizers operate on one-dimensional sequences of discrete tokens.
Single-cell expression data, by contrast, requires the simultaneous representation of gene identity (categorical), expression magnitude (continuous, sparse), and potentially multiple layers of metadata (hierarchical, categorical).
The \bmfmrna{} framework addresses this through a multifield tokenization system in which each cell is represented as an aligned set of parallel fields.

\subsubsection{Gene Encoding}
The \bmfmrna{} framework supports multiple gene representations: random initialization or pretrained embeddings from gene2vec~\cite{du2019gene2vec}, \esmTwo{}~\cite{lin2023evolutionary} (proteins), \DNABERTTwo{}~\cite{zhou2023dnabert} (DNA), and Borzoi~\cite{linder2025predicting} (epigenomics). Uniquely, we support partial coverage, learning random embeddings for non-coding genes alongside pretrained embeddings for coding genes.
For the downstream tasks that we have considered, the external embeddings did not provide improved results and were not utilized in our final pretrained models.
This finding is in line with recent ablation experiments indicating that with the standard projection architecture, external embeddings are not well utilized by TFMs~\cite{kalfon_scprint_2025}.

\subsubsection{Field structure}
For a cell expressing $k$ genes out of a vocabulary of $G$ total genes, the tokenizer produces a sequence of length $k$ (or $k+1$ when a \texttt{[CLS]} token is prepended), where each position carries:
(i)~a gene identity token, drawn from the gene vocabulary and mapped to either a randomly initialized or pretrained embedding;
(ii)~a continuous expression encoding, produced by the periodic or MLP-based encoder described in the main text (Section~\ref{p:cve}); and
(iii)~optional metadata tokens encoding per-gene annotations such as perturbation status.
Cell-level metadata (tissue, disease, donor) are encoded separately and injected via the \texttt{[CLS]} token or as additional conditioning signals.

\subsubsection{Handling of transcriptional zeros}
A fundamental challenge in single-cell tokenization is the treatment of zero-valued entries.
In a typical scRNA-seq profile, fewer than 3\% of genes have non-zero expression (median ${\sim}$1{,}700 out of $>$60{,}000 in the \cellxgene{} corpus), and it is generally not possible to distinguish true biological absence from technical dropout.
The framework supports multiple strategies for handling zeros:
(i)~\emph{exclusion}, in which only non-zero genes are included in the input sequence, reducing sequence length and computational cost at the expense of discarding potential biological signal;
(ii)~\emph{special token representation}, in which zero-valued genes are included and assigned a dedicated learned embedding, following Hao et al.~\cite{hao_scfoundation_2024}; and
(iii)~\emph{stochastic inclusion}, in which a random or stratified subset of zero-valued genes is appended to the non-zero set during each training step, providing regularization and partial exposure to silencing patterns.
For the WCED objective, zeros are excluded from the input but included in the reconstruction target, enabling the model to learn about gene silencing via the decoder without inflating input sequence length (see main text).

\subsubsection{Field-specific masking and dropout}
The multifield design enables masking and dropout strategies that operate independently on each field.
For example, during MLM pretraining the gene identity token may be replaced with a \texttt{[MASK]} token while the corresponding expression value is zeroed, or vice versa.
Field-specific dropout rates are configurable per experiment, allowing systematic ablation of input modalities.

\subsubsection{Input Representation}\label{s:input-representation}
The input to our model is designed to flexibly represent both gene identity and gene expression level, with the option of incorporating various levels of biological prior knowledge via pretrained embeddings. The encoder processes each cell as a structured sequence of gene-expression pairs, with embeddings constructed from multiple token types. This design supports both expressive modeling and downstream interpretability.

\paragraph{Mathematical Formulation}
scRNA-seq data provides a high-resolution view of gene expression at the single-cell level processed into cell-gene matrix.  Formally, let $M \in R_{NxG}$ be the cell-gene matrix, where $N\in {1, 2 ....,n}$ represent the number of cells, $G\in {1, 2 ....,g}$ represent the genes, then $M_{ng}$ represents the read count or expression of the $g^{th}$ gene in $n^{th}$ cell. The basic input to the \bmfmrna{} model is at a single cell consisting (1) genes in the cell, (2) gene expression profile, (3) gene-level states to represent metadata information associated with individual genes such as perturbation experiment states and (4) cell-level states to represent metadata information associated with individual cells such as cell type, tissue and disease contexts that are captured in large-scale single-cell transcriptomic datasets.

\paragraph{Expression Encoding}\label{p:cve}
In addition to the simple binning and tokenization approach used by \scBERT and some versions of \scGPT, \bmfmrna also supports continuous value encoding so that the full numerical information is available to the model in a learnable way. This is supported via two approaches:

\textit{Multi-layered Perceptron}: The MLP-based encoder transforms each expression value \( x_i \) through fully connected layers with non-linear activations:
\[
f_i(x_i) = W_n \cdot \phi(W_{n-1} \cdot \phi(\cdots \phi(W_1 x_i + b_1) \cdots) + b_{n-1}) + b_n
\]
where \( W_i \) are weight matrices, \( b_i \) are biases, and \( \phi(\cdot) \) is the activation function (ReLU or GELU).

\textit{Periodic activation}: Uses sine and cosine functions to capture periodic patterns:
\[
v_i = 2\pi c_i x_i \quad \text{and} \quad f_i(x_i) = \text{concat}[\sin(v_i), \cos(v_i)]
\]
where \( c_i \) is a trainable frequency parameter, followed by linear projection:
\[
f_{i, \text{final}} = W_2 \cdot \phi(W_1 f_i(x_i) + b_1) + b_2
\]

Additionally, the zero-special token mechanism treats zero expression as a distinct biological signal:
\[
\text{special\_token}(x_i) = \begin{cases}
\text{idx}_{\text{special}} & \text{if } x_i = 0 \\
x_i & \text{otherwise}
\end{cases}
\]
with embedding \( \text{special\_embedding}(x_i) = E_{\text{special}}(\text{idx}_{\text{special}}) \).

Based on our experimentation, the periodic encoding had lower loss on the same tasks, robust to its hyperparameters and that is what was used for the final pretrained checkpoints. We retained a dedicated token to represent the value 0, as in \cite{hao_scfoundation_2024}.

\subsection{Model Architectures}

\subsubsection{Architecture-agnostic design}
The framework separates the model backbone from input handling and output heads, so that any encoder architecture can be paired with any combination of tokenization strategy and training objective.
Architecture selection is specified in the Hydra configuration; switching between architectures requires no changes to data loading, training, or evaluation code.

\subsubsection{Supported architectures}
The following architectures are fully implemented and tested within the framework:

\paragraph{LLaMA (bidirectional)}
The primary architecture used in this work is a bidirectional adaptation of LLaMA~\cite{Touvron2023LLaMA}, with modifications for continuous biological inputs as described in the main text.
The key deviation from the standard LLaMA architecture is the removal of the pre-attention RMSNorm in the first transformer block, ensuring that the absolute magnitudes of continuous input embeddings are preserved before the first attention operation, following the design principle used in AlphaFold~3~\cite{abramson2024accurate} for coordinate injection.
Feed-forward layers use SwiGLU activations~\cite{Shazeer2020GLU}, and internal dropout is omitted in favor of weight decay regularization.

\paragraph{BERT and ModernBERT}
Standard \BERT{}~\cite{devlin2018bert} and \ModernBERT{} encoders are supported for compatibility with prior single-cell work and for ablation studies comparing architectural choices.

\paragraph{SCPerformer}
A Performer variant~\cite{choromanski2020rethinking} adapted for single-cell inputs, using random feature approximations to achieve linear-complexity attention.
This enables processing of full-transcriptome-length sequences without the quadratic memory cost of standard self-attention.

\emph{SCNystromformer.}
A Nystr\"omformer variant~\cite{xiong2021nystromformer} with custom iterative Moore--Penrose inverse approximations directly integrated into the attention mechanism.
Rather than relying on external library implementations, the framework provides a numerically stable iterative inverse computation that robustly handles the extreme context windows ($>$10{,}000 tokens) encountered in whole-transcriptome or long-read DNA sequence modeling.

\subsubsection{Output heads}
The framework implements a configurable multitask output head (\texttt{SCMultiTaskHead}) that routes representations from different positions in the encoder output to distinct prediction tasks.
Token-level embeddings are directed to per-gene reconstruction heads (for MLM or WCED), while the pooled \texttt{[CLS]} embedding is directed to sequence-level classification or regression heads.
Each head has independently configurable loss functions, loss weights, and activation schedules, enabling curriculum learning strategies in which objectives are introduced, removed, or reweighted over the course of training.

\subsection{Training Objectives}

The framework supports the following pretraining and fine-tuning objectives, which can be composed freely via configuration:

\subsubsection{Masked language modeling (MLM)}
A subset of input gene tokens is replaced with \texttt{[MASK]} tokens, and the model is trained to reconstruct the original gene identity and expression value from context.
Masking rates, masking strategies (uniform or adaptive), and reconstruction loss functions (cross-entropy for identity, MSE or zero-inflated MSE for expression) are independently configurable.

\subsubsection{Whole cell expression decoding (WCED)}
As described in the main text, the WCED objective trains the model to reconstruct the full expression profile of a cell from only the \texttt{[CLS]} token embedding, via a shallow decoder network.
The decoder produces two sets of outputs: binary detection logits (optimized with binary cross-entropy) and continuous expression predictions (optimized with MSE) for all genes in the vocabulary, not just those present in the input.
This forces the \texttt{[CLS]} embedding to serve as an information bottleneck encoding the complete cellular state.

\subsubsection{Hierarchical cell-type classification}
The hierarchical cross-entropy loss described in the main text (Section~\ref{sec:hierarchical_loss}) enables ontology-aware supervision during pretraining.
Ground-truth labels at any depth in the Cell Ontology DAG are accommodated by computing the predicted probability mass over the subtree of descendant leaves.

\subsubsection{Gradient reversal for batch correction}
A gradient reversal layer (GRL) can be attached to the \texttt{[CLS]} embedding to adversarially penalize the encoding of batch-specific information, encouraging the model to learn batch-invariant representations during pretraining.

\subsection{Evaluation and Interpretability}

\subsubsection{Evaluation suite}
The repository includes standardized evaluation protocols for the tasks reported in the main text, including zero-shot batch integration (via scib metrics~\cite{luecken_benchmarking_2022}), zero-shot and fine-tuned cell-type annotation (via F1, accuracy, and hierarchical accuracy), and embedding quality assessment (via the CZ Benchmark~\cite{cz-benchmarks}).
Evaluation datasets, preprocessing steps, data splits, and metric computations are specified in Hydra configurations and executed through a unified evaluation entry point, ensuring that all reported numbers are reproducible from a single command.

\subsubsection{Gene-level error tracking}
During both pretraining and evaluation, the framework records per-gene reconstruction errors (detection AUC and expression MAE) at every validation epoch.
These gene-level error trajectories enable the learnability analyses presented in the main text and provide a diagnostic tool for understanding what biological signals the model captures at different stages of training.

\subsubsection{Interpretability via integrated gradients}
The framework integrates Captum~\cite{kokhlikyan2020captum} to compute layer integrated gradients for any model prediction.
Given a trained model and a target output (e.g., a cell-type classification logit), the method attributes the prediction to individual input gene tokens, producing a per-gene importance score.
This enables systematic identification of the genes driving the model's predictions for specific cell types or conditions.

\subsection{Multi-Omic Extensibility}

Although this manuscript demonstrates the transcriptomic (\bmfmrna{}) instantiation of the framework, the underlying software was designed for multi-omic generalization.
The repository includes the following additional capabilities, which are implemented and functional but not the focus of the present evaluation:

\subsubsection{DNA and epigenomic modeling}
Dedicated data modules and tokenization schemes for DNA sequences, including \texttt{snp2vec} (encoding single-nucleotide polymorphisms as learned embeddings), \texttt{ref2vec} (encoding reference genome context), and hierarchical chunking strategies for long genomic intervals.
These modules support variant-to-trait prediction, chromatin accessibility modeling, and regulatory element classification tasks.

\paragraph{Perturbation biology.}
Preprocessing pipelines and data modules for large-scale perturbation screens, including the Replogle~\cite{replogle_mapping_2022} and Norman~\cite{norman_exploring_2019} CRISPR datasets.
These modules standardize the loading, quality control, and train/test splitting of perturbation data, and interface with the same multitask training infrastructure used for transcriptomic pretraining.

\paragraph{Cross-modal architecture.}
Because the multifield tokenization and multitask head systems are agnostic to the biological modality of the input, the same training infrastructure can in principle accommodate joint modeling of transcriptomic, genomic, and epigenomic data within a single model, provided appropriate tokenizers and loss functions are configured.

\section{Datasets used}
\label{sec:datasets}

\begin{table}[htbp]
\centering
\caption{Evaluation datasets used in this study. Cell counts reflect processed versions used for evaluation; original datasets may contain additional cells removed during quality control. Datasets marked with $\dagger$ are from the scIB benchmark~\cite{luecken_benchmarking_2022}; those marked with $\ddagger$ are from the scEval benchmark~\cite{liu_evaluating_2024}; those marked with $\star$ use scGPT-shared splits~\cite{cui_scgpt_2024}.}
\label{tab:datasets}
\tiny
\setlength{\tabcolsep}{3pt}
\begin{tabular}{@{}l@{\hspace{0.2cm}}l@{\hspace{0.2cm}}l@{\hspace{0.2cm}}l@{\hspace{0.2cm}}r@{\hspace{0.2cm}}l@{\hspace{0.2cm}}l@{\hspace{0.2cm}}l@{}}
\toprule
\textbf{Dataset} & \textbf{Accession} & \textbf{Study} & \textbf{Tissue} & \textbf{Cells} & \textbf{Species} & \textbf{Platform} & \textbf{Task(s)} \\
\midrule
\multicolumn{8}{l}{\textit{Zero-shot batch integration and cell type annotation (scEval, human)}} \\
\addlinespace[2pt]
\texttt{cell\_lines}$^{\dagger\ddagger}$ & GSE118767 & Tian et al.~\cite{tian_benchmarking_2019} & Cell lines & 1,916 & Human & 10X/CEL-seq2/Drop-seq & BI, CT \\
\texttt{covid\_19}$^{\ddagger}$ & E-MTAB-10026$^{f}$ & Stephenson et al.~\cite{stephenson_single-cell_2021} & PBMCs & 780K+ & Human & 10X 5$'$ CITE-seq & BI, CT \\
\texttt{dc}$^{\dagger\ddagger}$ & GSE94820 & Villani et al.~\cite{villani_single-cell_2017} & Blood DCs & 2,400 & Human & Smart-seq2 & BI, CT \\
\texttt{hbones}$^{\ddagger}$ & GSE120221 & Oetjen et al.~\cite{oetjen_human_2018} & Bone marrow & 30K+ & Human & 10X Chromium & BI, CT \\
\texttt{heart\_atlas}$^{\ddagger}$ & ERP123138 & Litvi\v{n}ukov\'{a} et al.~\cite{litvinukova_cells_2020} & Heart & 487K & Human & 10X v2/v3 & BI, CT \\
\texttt{human\_pbmc}$^{\dagger\ddagger}$ & GSE132044 & Ding et al.~\cite{ding_systematic_2020} & PBMCs & 33K & Human & Multiple (7 methods) & BI, CT \\
\texttt{immune\_all\_human}$^{\dagger\ddagger}$ & Multiple$^{a}$ & Luecken et al.~\cite{luecken_benchmarking_2022} & Blood/BM & 33K & Human & Multiple & BI, CT \\
\texttt{immune\_atlas}$^{\dagger\ddagger}$ & figshare$^{b}$ & Luecken et al.~\cite{luecken_benchmarking_2022} & Immune & 98K & Hu./Mu. & Multiple & BI, CT \\
\texttt{lung\_atlas}$^{\dagger\ddagger}$ & figshare$^{b}$ & Vieira Braga et al.~\cite{vieira_braga_cellular_2019} & Lung & 32K & Human & Drop-seq/10X & BI, CT \\
\texttt{pancrm}$^{\dagger\ddagger}$ & Multiple$^{c}$ & Baron, Muraro et al.$^{c}$ & Pancreas & 16K & Human & Multiple (5 protocols) & BI, CT \\
\texttt{pbmc\_10k}$^{\ddagger}$ & 10X Genomics & 10X Genomics~\cite{10x_pbmc10k_2018} & PBMCs & 12K & Human & 10X 3$'$ v3 & BI, CT \\
\texttt{zheng68k}$^{\ddagger}$ & SRP073767 & Zheng et al.~\cite{zheng_massively_2017} & PBMCs & 68K & Human & 10X Chromium & CT \\
\addlinespace[4pt]
\multicolumn{8}{l}{\textit{Zero-shot batch integration (scEval, mouse)}} \\
\addlinespace[2pt]
\texttt{mca}$^{\ddagger}$ & GSE108097 & Han et al.~\cite{han_mapping_2018} & Multiple (mouse) & 400K+ & Mouse & Microwell-seq & BI, CT \\
\texttt{mhsp}$^{\ddagger}$ & GSE137539 & Zhu et al.~\cite{zhu_developmental_2020} & Hemato.\ prog. & 5K & Mouse & 10X Chromium & BI, CT \\
\addlinespace[4pt]
\multicolumn{8}{l}{\textit{Cell type annotation}} \\
\addlinespace[2pt]
\texttt{pancreascross}$^{\ddagger}$ & Multiple$^{c}$ & Baron, Muraro et al.$^{c}$ & Pancreas & 16K & Human & Multiple & CT \\
\texttt{multiple\_sclerosis}$^{\star}$ & E-HCAD-35 & Schirmer et al.~\cite{schirmer_neuronal_2019} & Brain & 8K & Human & 10X snRNA-seq & CT (OOD) \\
\texttt{myeloid}$^{\star}$ & GSE154763 & Cheng et al.~\cite{cheng_pan-cancer_2021} & Pan-cancer myeloid & 55K & Human & 10X Chromium & CT \\
\addlinespace[4pt]
\multicolumn{8}{l}{\textit{CZI Benchmark cell type classification (Tabula Sapiens v2)}} \\
\addlinespace[2pt]
TS v2 Blood & CELLxGENE$^{d}$ & TS Consortium~\cite{tabula_sapiens_2022,tabula_sapiens_v2} & Blood & \multirow{5}{*}{500K+ total} & Human & 10X/Smart-seq2 & CT \\
TS v2 Bone Marrow & CELLxGENE$^{d}$ & TS Consortium~\cite{tabula_sapiens_2022,tabula_sapiens_v2} & Bone marrow & & Human & 10X/Smart-seq2 & CT \\
TS v2 Lung & CELLxGENE$^{d}$ & TS Consortium~\cite{tabula_sapiens_2022,tabula_sapiens_v2} & Lung & & Human & 10X/Smart-seq2 & CT \\
TS v2 Mammary & CELLxGENE$^{d}$ & TS Consortium~\cite{tabula_sapiens_2022,tabula_sapiens_v2} & Mammary & & Human & 10X/Smart-seq2 & CT \\
TS v2 Thymus & CELLxGENE$^{d}$ & TS Consortium~\cite{tabula_sapiens_2022,tabula_sapiens_v2} & Thymus & & Human & 10X/Smart-seq2 & CT \\
\addlinespace[4pt]
\multicolumn{8}{l}{\textit{Disease case study}} \\
\addlinespace[2pt]
Alzheimer's & GSE138852 & Grubman et al.~\cite{grubman_single-cell_2019} & Entorhinal cortex & 13,214 & Human & 10X / NextSeq 500 & CS \\
\addlinespace[4pt]
\multicolumn{8}{l}{\textit{Pretraining corpus}} \\
\addlinespace[2pt]
CELLxGENE Census & TileDB-SOMA & CZI~\cite{program_cz_2025} & Multi-tissue & 6.2M$^{e}$ & Human & Multiple (31 assays) & PT \\
\bottomrule
\end{tabular}
\vspace{4pt}

{\footnotesize
BI: batch integration; CT: cell type annotation; PP: perturbation prediction; CS: case study; PT: pretraining; OOD: out-of-distribution.\\
$^{a}$~Compiled from GSE120221, 10X PBMC datasets, Freytag et al., Sun et al., and Villani et al.; see Luecken et al.~\cite{luecken_benchmarking_2022}.\\
$^{b}$~scIB figshare: \url{https://doi.org/10.6084/m9.figshare.12420968}.\\
$^{c}$~Five pancreas studies: Baron et al.~\cite{baron_single-cell_2016} (GSE84133), Muraro et al.~\cite{muraro_single-cell_2016} (GSE85241), Segerstolpe et al.~\cite{segerstolpe_single-cell_2016} (E-MTAB-5061), Lawlor et al.~\cite{lawlor_single-cell_2017} (GSE86469), Gr\"{u}n et al.~\cite{grun_novo_2016} (GSE81076).\\
$^{d}$~CZ Benchmarks~\cite{cz-benchmarks}; v2-only benchmark excludes v1 samples.\\
$^{e}$~10\% downsample of the 2025-01-30 LTS release ($\approx$62M human cells total); 19,362 protein-coding genes per HGNC.
} \\
$^{f}$Covid-19 dataset was excluded from cell type annotation task evaluation since its two batches have largely non-overlapping cell type compositions.
\end{table}